\documentclass{article}
\usepackage{arxiv}
\usepackage{graphicx}
\graphicspath{{Figures/}}
\usepackage{subfigure}
\usepackage{amsmath,amsfonts,amssymb}
\usepackage{lineno}
\usepackage{siunitx}
\usepackage{color}

\usepackage[utf8]{inputenc}
\usepackage[numbers]{natbib}
\usepackage{graphicx}
\providecommand{\keywords}[1]{\textbf{\textit{Index terms---}} #1}

\newcommand{\transpose}{^\intercal}

\title{Latent-space time evolution of non-intrusive reduced-order models using Gaussian process emulation}
\author{
  Romit Maulik \\
  Argonne Leadership Computing Facility \\
  Argonne National Laboratory \\
  Lemont, IL 60439 \\
  \texttt{rmaulik@anl.gov} \\
  \And
  Themistoklis Botsas \\
  The Alan Turing Institute \\
  London NW1 2DB \\
  \texttt{tbotsas@turing.ac.uk}
  \And
  Nesar Ramachandra \\
  High Energy Physics Division \\
  Argonne National Laboratory \\
  Lemont, IL 60439 \\
  \texttt{nramachandra@anl.gov}
  \And
  Lachlan Robert Mason \\
  Imperial College London SW7 2AZ \\
  The Alan Turing Institute, NW1 2DB \\
  \texttt{lmason@turing.ac.uk}
  \And
  Indranil Pan \\
  Imperial College London SW7 2AZ \\
  The Alan Turing Institute, NW1 2DB\\
  \texttt{i.pan11@imperial.ac.uk}
}
\date{}

\begin{document}

\maketitle

\begin{abstract}
    Non-intrusive reduced-order models (ROMs) have recently generated considerable interest for constructing computationally efficient counterparts of nonlinear dynamical systems emerging from various domain sciences. They provide a low-dimensional emulation framework for systems that may be intrinsically high-dimensional. This is accomplished by utilizing a construction algorithm that is purely data-driven. It is no surprise, therefore, that the algorithmic advances of machine learning have led to non-intrusive ROMs with greater accuracy and computational gains. However, in bypassing the utilization of an equation-based evolution, it is often seen that the interpretability of the ROM framework suffers. This becomes more problematic when black-box deep learning methods are used which are notorious for lacking robustness outside the physical regime of the observed data. In this article, we propose the use of a novel latent-space interpolation algorithm based on Gaussian process regression. Notably, this reduced-order evolution of the system is parameterized by control parameters to allow for interpolation in space. The use of this procedure also allows for a continuous interpretation of time which allows for temporal interpolation. The latter aspect provides information, with quantified uncertainty, about full-state evolution at a finer resolution than that utilized for training the ROMs. We assess the viability of this algorithm for an advection-dominated system given by the inviscid shallow water equations.
\end{abstract}

\keywords{Reduced-order models; Deep learning; Gaussian process regression}

\section{Introduction}

Recently, researchers have shown sustained interest in using machine learning (ML) methods for non-intrusive reduced-order models (ROMs) for systems that may be governed by advection-dominated partial differential equations (PDEs). This is because solving PDE forward-models for such systems may require very fine spatiotemporal numerical discretizations which cause a significant bottleneck in design and forecast tasks \cite{verstappen1997direct}. The prospect of bypassing traditional numerical methods and building surrogates from data alone \cite{long2017pde,raissi2018deep,raissi2019physics,sirignano2018dgm} is attractive for multiple applications ranging from engineering design \cite{bui2004aerodynamic,renganathan2018koopman} and control \cite{noack2011reduced,kutz2016dynamic} to climate modeling \cite{chattopadhyay2020predicting,chattopadhyay2019analog,maulik2020recurrent}. This is because data-driven ROMs allow for rapid predictions of nonlinear dynamics unencumbered by the stability-based limitations of numerical discretizations. In almost all ROM applications, forecasts must be conditioned on time and several control parameters, such as the initial conditions or the physical properties of the governing laws. In addition, since these models eschew the use of PDEs, it is necessary to associate some notion of uncertainty quantification during the time evolution of these surrogate dynamical systems. This is to ensure that the loss of interpretability and reliability by bypassing equations is offset by a feedback process from the ML.

Neural networks have been used for ROMs for decades. One of the earliest examples \cite{hsieh1998applying} used a simple fully connected network for forecasting meteorological information.  More recently, researchers  have incorporated a single-layered feed-forward neural network into a nonlinear dynamical system and built a surrogate model for a high-dimensional aerodynamics problem \cite{mannarino2014nonlinear}; radial basis function networks have been used to make forecasts for a nonlinear unsteady aerodynamics task \cite{zhang2016nonlinear,kou2017layered}; and a simple fully connected network has been used for learning the dynamics of an advection-dominated system \cite{san2018neural,hesthaven2018non}. 

Neural networks are commonly used for two tasks in typical ROM construction: Compression and time-evolution. For the former, they may be used as a nonlinear equivalent of the proper orthogonal decomposition (POD) or principal component analysis (PCA) based methods which find a linear affine subspace for the full system. The identification of this reduced basis to ensure a compressed representation that is minimally \emph{lossy} is a core component of most ROM development strategies (some examples include \cite{san2014basis,korda2018data,kalb2007intrinsic}). While POD-based methods currently represent the most popular technique for reduced-basis (or latent space) construction, data generated from PDE simulations can often be interpreted as images on a square grid; therefore, convolutional neural networks have also been applied \cite{thuerey2020deep,kim2019deep,gonzalez2018learning,maulik2020time} for building computationally effective ROMs.

Once this basis (or latent representation) is identified, we need a cost-effective strategy for accurate nonlinear dynamical system evolution to reproduce the full-order spatiotemporal complexity of the problem in the reduced basis. For example, linear reduced-basis construction allows for the use of intrusive methods (which project the governing equations onto the reduced-basis), as seen in \cite{kalashnikova2010stability,mohebujjaman2019physically}, which use a Galerkin projection; or \cite{carlberg2011efficient,xiao2013non,fang2013non}, which use the Petrov–Galerkin method. Such extensions are not straightforward for autoencoder-based latent-space constructions, since projecting to a basis space is infeasible. We note, though, that the use of convolutional autoencoder architectures has been demonstrated for the Petrov–Galerkin method, where there is no requirement to project the governing equations onto a trial space \cite{lee2020model}. As mentioned previously, neural networks are also commonly used for the temporal evolution of these latent-space representations. Mainstream ML has generated a large body of time-series forecasting literature that lends itself readily to latent-space evolution of dynamical systems. Recently, long short-term memory architectures (LSTMs) have become popular for the non-intrusive characterization of dynamical systems \cite{vlachas2018data,ahmed2020long,mohan2019compressed,mohan2018deep,wang2019recurrent} as they allow for the construction of non-independent and non-identically distributed, directional relationships between the states of a dynamical system at different times before a future forecast is performed. Other methods that have been utilized for such tasks include the temporal convolutional network \cite{xu2019multi}, non-autoregressive variants for the LSTM \cite{maulik2020non} and system-identification within latent space \cite{champion2019data}. In most of these developments, the evolution framework (in time) is \emph{deterministic} in nature. These frameworks highlight a crucial drawback when coupled with the black-box nature of purely data-driven ROMs – it is imperative to embed some notion of uncertainty into the time evolution of ROMs in the latent space. Past work has shown that LSTMs suffer from stability and interpretability issues when utilized for the reduced-order modeling of advection-dominated systems \cite{maulik2020non} and this study proposes an alternative to provide interpretable forecasts with quantified uncertainty. To that end, we propose the use of a Gaussian process regression (GPR) framework for evolving the low-dimensional representation of data obtained from a PDE evolution. \textcolor{black}{A Gaussian process (GP) is an infinite-dimensional generalization of a multivariate normal distribution. Instead of assigning a specific function and estimating parameters to fit training data, the GPR assigns a probability over the function space. Such a probabilistic approach not only allows for the most probable predictions (mean of the distribution) but also for characterizing the uncertainty in the predictions themselves.} In addition to providing forecasts at the temporal resolution of the training data, the use of the framework allows for interpolation in time. This is because time is interpreted as a continuous variable. When coupled with quantified uncertainty, the framework allows a ROM user to interrogate the behavior of the emulated system at a finer temporal resolution, with implications for data sources that are temporally sparse. 

To summarize, the major contributions of this article are:
\begin{itemize}
    \item We introduce a technique to construct parametric non-intrusive ROMs with quantified uncertainty during latent-space time evolution of dynamical systems.
    \item We detail the use of Gaussian processes (GPs) that are customized for the time-evolution of parametric nonlinear dynamical systems.
    \item We demonstrate the ability of the proposed time-evolution algorithm for systems compressed by linear reduced-basis methods such as POD, as well as nonlinear compression frameworks such as variational and convolutional autoencoders.
    \item We test the ability of a ROM constructed from coarse training data to interpolate on a finer temporal resolution with quantified uncertainty.
\end{itemize}

In the following, we shall introduce our experimental setup which relies on the inviscid shallow water equations in Section 2, our various compression mechanisms in Section 3, the UQ embedded emulation strategy in Section 4, followed by results and conclusions in Sections 5 and 6 respectively.

\begin{figure}
    \centering{\includegraphics[width=\textwidth]{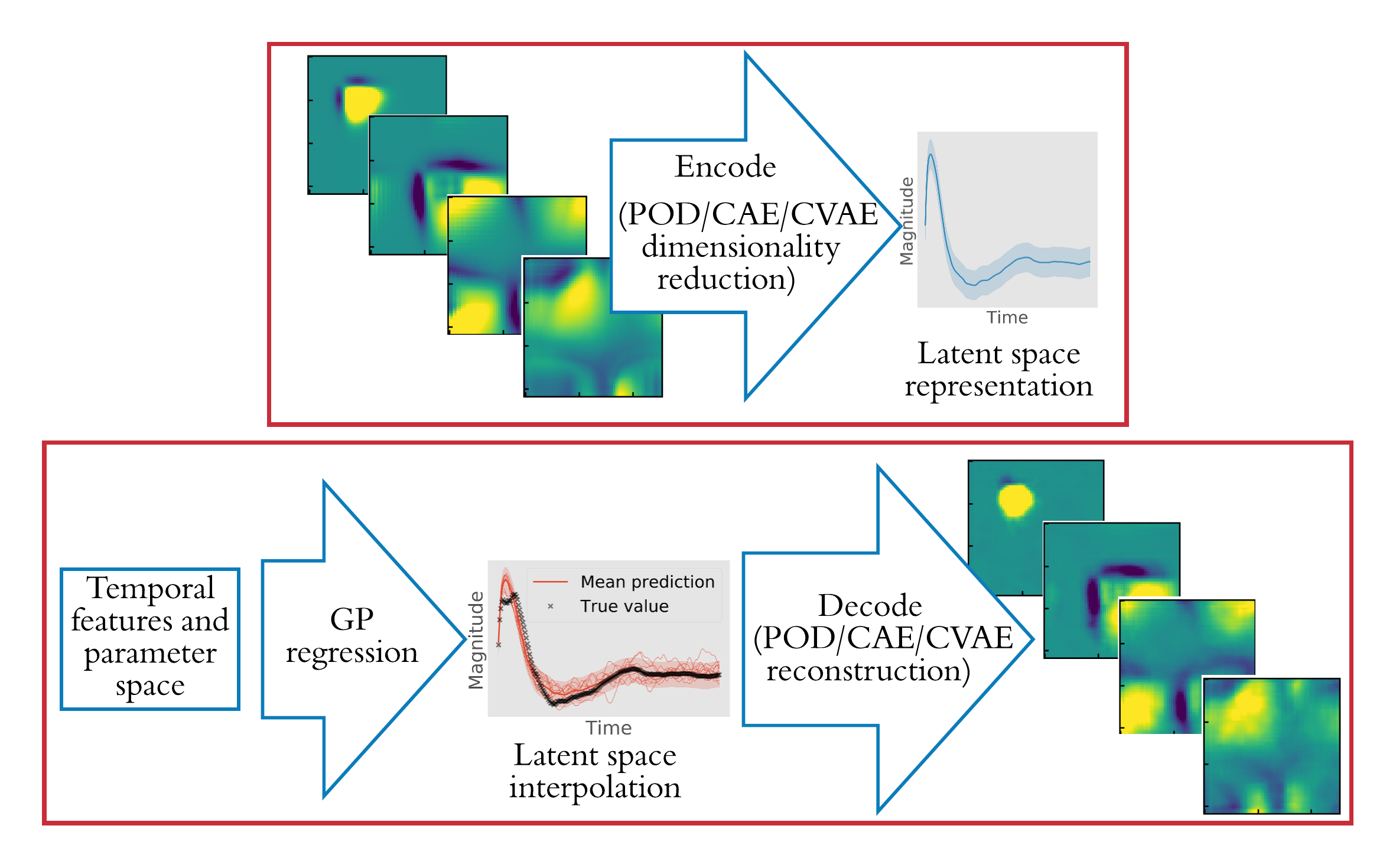}}
    \caption{Schematic of reduced-order modeling. Upper panel shows the dimensionality reduction operation to a latent space. Lower panel shows the forecasting in the latent space for reconstructions.}
    \label{schematic}
\end{figure}

\section{The shallow water equations}

The inviscid shallow water equations belong to a prototypical system of equations for geophysical flows \cite{kalnay2003atmospheric}. In particular, the shallow water equations admit solutions where advection dominates dissipation and pose challenges for conventional ROMs \cite{wang2012proper}. These governing equations are given by
\begin{align}
    \frac{\partial(\rho \eta)}{\partial t}+\frac{\partial(\rho \eta u)}{\partial x}+\frac{\partial(\rho \eta v)}{\partial y} =0  \label{eq1} \\
    \frac{\partial(\rho \eta u)}{\partial t}+\frac{\partial}{\partial x}\left(\rho \eta u^{2}+\frac{1}{2} \rho g \eta^{2}\right)+\frac{\partial(\rho \eta u v)}{\partial y} = 0 \label{eq2} \\
    \frac{\partial(\rho \eta v)}{\partial t}+\frac{\partial(\rho \eta u v)}{\partial x}+\frac{\partial}{\partial y}\left(\rho \eta v^{2}+\frac{1}{2} \rho g \eta^{2}\right) = 0, \label{eq3}
\end{align}
where, $\eta$ corresponds to the total fluid column height, and $(u,v)$ is the fluid's horizontal flow velocity, averaged across the vertical column, $g$ is acceleration due to gravity, and $\rho$ is the fluid density, typically set to \num{1.0}. Here, $t$, $x$ and $y$ are the independent variables of time and the spatial coordinates of the two-dimensional system. Equation \ref{eq1} captures the law of mass conservation, whereas Equations \ref{eq2} and \ref{eq3} denote the conservation of momentum. The initial conditions of the problem are given by 
\begin{align}
    \rho \eta (x,y,t=0) &= 1 + \exp\left [-\left(\frac{(x-\bar{x})^2}{2(\num{5e+4})^2} + \frac{(y-\bar{y})^2}{2(\num{5e+4})^2}\right) \right] \\
    \rho \eta u(x,y,t=0) &= 0 \\
    \rho \eta v(x,y,t=0) &= 0,
\end{align}
i.e., a Gaussian perturbation at a particular location on the grid $[\bar{x},\bar{y}] \equiv w$. We solve the system of equations until $t=0.5$ with a time-step of \num{0.001} seconds on a square two-dimensional grid with \num{64} collocation points to completely capture the advection and gradual decay of this perturbation. Note that these numbers may vary according to the forecasting and fidelity requirements of a particular problem and perturbation. The initial and boundary conditions for this particular shallow-water equation experiment represent a tightly-controlled traveling wave problem that is translation invariant. Different realizations of the initial condition lead to translationally shifted trajectories. We also note the presence of mirror symmetries with respect to $x=\bar{x}$ and $y=\bar{y}$ coupled with a rotational symmetry of $\pi/2$ radians about the origin. However, our motivation for a first assessment of our emulators on this system stems from the well-known fact that POD and Galerkin-projection based methods are severely limited in their ability to forecast on these simple traveling-wave systems \cite{KFB2016,mendible2019dimensionality} and require special treatment with intrinsic knowledge of the flow dynamics. This is in addition to the fact that equation-based models are impossible to construct because of the absence of information from the other variables of the PDE. We seek to build predictive models solely from observations of $\rho \eta$ conditioned on $w$ mimicking a real-world scenario where complete observations of all relevant variables (in this case, velocities) are unavailable. 

\section{Data-compression}\label{sec:data_compression}

\subsection{Proper orthogonal decomposition}

In this section, we review the POD technique for the construction of a reduced basis \cite{kosambi1943statistics,berkooz1993proper}. The interested reader may also find an excellent explanation of POD and its relationship with other dimension-reduction techniques in \cite{taira2019modal}. The POD procedure is tasked with identifying a space
\begin{linenomath*}
\begin{align}
\mathbf{X}^{f}=\operatorname{span}\left\{\boldsymbol{\vartheta}^{1}, \dots, \boldsymbol{\vartheta}^{f}\right\},
\end{align}
\end{linenomath*}
which approximates snapshots optimally with respect to the $L^2$-norm. The process of $\boldsymbol{\vartheta}$ generation commences with the collection of snapshots in the \emph{snapshot matrix}
\begin{linenomath*}
\begin{align}
\mathbf{S} = [\begin{array}{c|c|c|c}{\hat{\mathbf{q}}^{1}_h} & {\hat{\mathbf{q}}^{2}_h} & {\cdots} & {\hat{\mathbf{q}}^{N_{s}}_h}\end{array}] \in \mathbb{R}^{N_{h} \times N_{s}},
\end{align}
\end{linenomath*}
where $N_s$ is the number of snapshots, and $\hat{\mathbf{q}}^i_h : \mathcal{T} \times \mathcal{P} \rightarrow \mathbb{R}^{N_h}$ corresponds to an individual snapshot in time of the discrete solution domain with the mean value removed, i.e.,
\begin{linenomath*}
\begin{align}
\begin{gathered}
\hat{\mathbf{q}}^i_h = \mathbf{q}^i_h - \mathbf{\bar{q}}_h, \\
\mathbf{\bar{q}}_h = \frac{1}{N_s} \sum_{i=1}^{N_s} \mathbf{q}^i_h.
\end{gathered}
\end{align}
\end{linenomath*}
with $\overline{\mathbf{q}}_h : \mathcal{P} \rightarrow \mathbb{R}^{N_h}$ being the time-averaged solution field. Our POD bases can then be extracted efficiently through the method of snapshots where we solve the eigenvalue problem on the correlation matrix $\mathbf{C} = \mathbf{S}\transpose \mathbf{S} \in \mathbb{R}^{N_s \times N_s}$. Then
\begin{linenomath*}
\begin{align}
\begin{gathered}
\mathbf{C} \mathbf{W} = \mathbf{W} \Lambda,
\end{gathered}
\end{align}
\end{linenomath*}
where $\Lambda = \operatorname{diag}\left\{\lambda_{1}, \lambda_{2}, \cdots, \lambda_{N_{s}}\right\} \in \mathbb{R}^{N_{s} \times N_{s}}$ is the diagonal matrix of eigenvalues and $\mathbf{W} \in \mathbb{R}^{N_{s} \times N_{s}}$ is the eigenvector matrix. Our POD basis matrix can then be obtained by
\begin{linenomath*}
\begin{align}
\begin{gathered}
\boldsymbol{\vartheta} = \mathbf{S} \mathbf{W} \in \mathbb{R}^{N_h \times N_s}.
\end{gathered}
\end{align}
\end{linenomath*}
In practice a reduced basis $\boldsymbol{\psi} \in \mathbb{R}^{N_h \times N_r}$ is built by choosing the first $N_r$ columns of $\boldsymbol{\vartheta}$ for the purpose of efficient ROMs, where $N_r \ll N_s$. This reduced basis spans a space given by
\begin{linenomath*}
\begin{align}
\mathbf{X}^{r}=\operatorname{span}\left\{\boldsymbol{\psi}^{1}, \dots, \boldsymbol{\psi}^{N_r}\right\}.
\end{align}
\end{linenomath*}
The coefficients of this reduced basis (which capture the underlying temporal effects) may be extracted as
\begin{linenomath*}
\begin{align}
\begin{gathered}
\mathbf{A} = \boldsymbol{\psi}\transpose \mathbf{S} \in \mathbb{R}^{N_r \times N_s}.
\end{gathered}
\end{align}
\end{linenomath*}
The POD approximation of our solution is then obtained via
\begin{linenomath*}
\begin{align}
\hat{\mathbf{S}} =  [\begin{array}{c|c|c|c}{\tilde{\mathbf{q}}^{1}_h} & {\tilde{\mathbf{q}}^{2}_h} & {\cdots} & {\tilde{\mathbf{q}}^{N_{s}}_h}\end{array}] \approx \boldsymbol{\psi} \mathbf{A} \in \mathbb{R}^{N_h \times N_s},
\end{align}
\end{linenomath*}
where $\tilde{\mathbf{q}}_h^i : \mathcal{T} \times \mathcal{P} \rightarrow \mathbb{R}^{N_h}$ corresponds to the POD approximation to $\hat{\mathbf{q}}_h^i$. The optimal nature of reconstruction may be understood by defining the relative projection error
\begin{linenomath*}
\begin{align}
\frac{\sum_{i=1}^{N_{s}}\left\|\hat{\mathbf{q}}^i_h-\tilde{\mathbf{q}}^i_h \right\|_{\mathbb{R}^{N_{h}}}^{2}}{\sum_{i=1}^{N_{s}}\left\|\hat{\mathbf{q}}^i_h\right\|_{\mathbb{R}^{N_{h}}}^{2}}=\frac{\sum_{i=N_r+1}^{N_{s}} \lambda_{i}^{2}}{\sum_{i=1}^{N_{s}} \lambda_{i}^{2}},
\end{align}
\end{linenomath*}
which exhibits that with increasing retention of POD bases, increasing reconstruction accuracy may be obtained. We remark that for dimension $d>1$, the solution variables may be stacked to obtain this set of bases that are utilized for the reduction of each PDE within the coupled system. Another approach may be to obtain reduced bases for each dependent variable within the coupled system and evolve each PDE on a different manifold. Each dependent variable is projected onto bases constructed from its snapshots alone. This affects the computation of the nonlinear term for computing the updates for each dimension in $\mathbf{q}$. In practice, this operation manifests itself in the concatenation of reduced bases to obtain one linear operation for reconstruction of all field quantities.

\subsection{Convolutional autoencoders}

\begin{figure}
    \centering{\includegraphics[width=\textwidth]{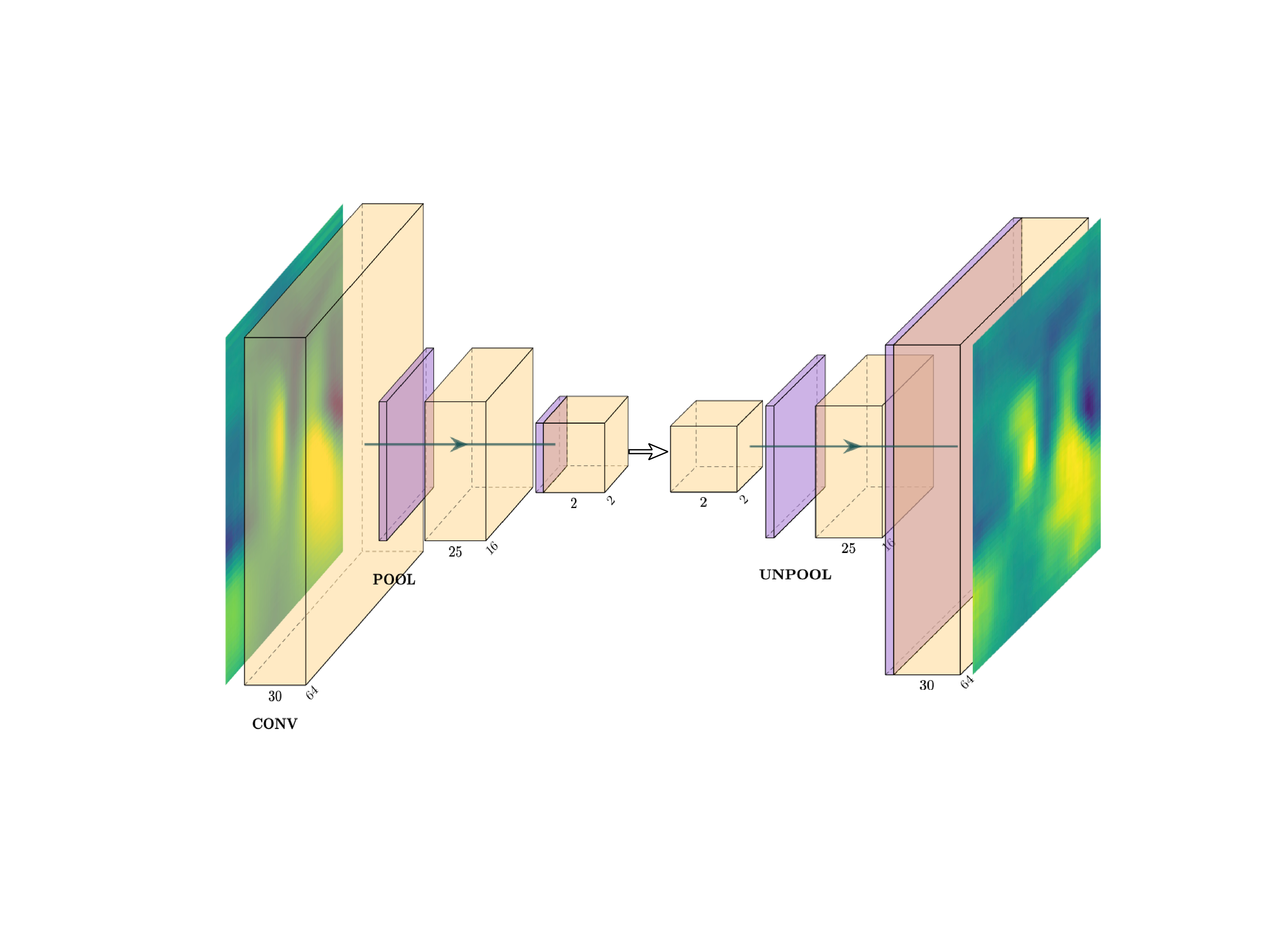}}
    \caption{Network architecture shows the encoder, bottleneck and decoder of a typical convolutional autoencoder (CAE). The input image is reconstructed using convolution, pooling and upsampling/unpooling layers.}
    \label{cae_arch}
\end{figure}

Autoencoders are neural networks that learn a new representation of the input data, usually with lower dimensionality. The initial layers, called the \emph{encoder}, map the input $\mathbf{x}\in \mathbb{R}^m$ to a new representation $\mathbf{z} \in \mathbb{R}^k$. The remaining layers, called the \emph{decoder}, map $\mathbf{z}$ back to $\mathbb{R}^m$ with the goal of reconstructing $\mathbf{x}$. The objective is to minimize the reconstruction error. Autoencoders are unsupervised; the data $\mathbf{x}$ is given, but the representation $\mathbf{z}$ must be learned.

More specifically, we use convolutional autoencoders (CAEs) that have convolutional layers\textcolor{black}{, shown in Figure \ref{cae_arch}}. In a convolutional layer, instead of learning a matrix that connects all $m$ neurons of layer's input to all $n$ neurons of the layer's output, we learn a set of filters. Each filter $\mathbf{f}_i$ is convolved with patches of the layer's input. Suppose a one-dimensional convolutional layer has filters of length $m_{\mathbf{f}_i}$. Then each of the layer's output neurons corresponding to filter $\mathbf{f}_i$ is connected to a patch of $m_{\mathbf{f}_i}$ of the layer's input neurons. In particular, a one-dimensional convolution of filter $\mathbf{f}$ and patch $\mathbf{p}$ is defined as $\mathbf{f} \ast \mathbf{p} = \sum_j f_j p_j$ (for neural networks, convolutions are usually technically implemented as cross-correlations). Then, for a typical one-dimensional convolutional layer, the layer's output neuron $y_{ij} = \varphi (\mathbf{f}_i \ast \mathbf{p}_j +B_{i})$, where $\varphi$ is an activation function and $B_i$ are the entries of a bias term. As $j$ increases, patches are shifted by stride $s$. For example, a one-dimensional convolutional layer with a filter $\mathbf{f}_0$ of length $m_{\mathbf{f}_0} = 3$ and stride $s=1$ could be defined so that $y_{0j}$ involves the convolution of $\mathbf{f}_0$ and inputs $j-1, j$, and $j+1$. To calculate the convolution, it is common to add zeros around the inputs to a layer, which is called \emph{zero padding}. 

Two-dimensional convolutions are defined similarly, but each filter and each patch are two-dimensional. A two-dimensional convolution sums over both dimensions, and patches are shifted both ways. For a typical two-dimensional convolutional layer, the output neuron $y_{ijk} = \varphi (\mathbf{f}_i \ast \mathbf{p}_{jk} +B_{i})$. Input data can also have a ``channel'' dimension, such as red, green and blue for images. The convolutional operator sums over channel dimensions, but each patch contains all of the channels. The filters remain the same size as patches, so they can have different weights for different channels. In our study, we use solely one channel for the spatial magnitude of $\rho \eta$. \textcolor{black}{These convolution layers, shown as yellow blocks in Figure \ref{cae_arch}, provide two-dimensional filter outputs essentially forming a set of feature maps for the input image. These convolutional layers are stacked, and more abstract features are captured within deeper layers of the encoder.}

It is common to follow a convolutional layer with a \emph{pooling} layer, which outputs a sub-sampled version of the input. In this paper, we specifically use max-pooling layers. Each output of a max-pooling layer is connected to a patch of the input, and it returns the maximum value in the patch. \textcolor{black}{The pooling layer thus summarizes a version of the features detected in the input. Stacking the pool layers along with convolution layers in the encoder thus provides a compact representation of the data at the bottleneck of the autoencoder, at center of the Figure \ref{cae_arch}.} 

\textcolor{black}{The decoder, (right half of Figure \ref{cae_arch}) is tasked with reconstructing the image from the latent-space representation at the bottleneck. This involves stacking deconvolutional layers (which perform a transposed convolution operation) to return to the original dimension. These layers upsample/unpool with nearest-neighbor interpolation. The loss function of the CAE is calculated by computing the difference between the output of the the final layer of the decoder and the input training image. Minimizing this reconstruction loss function in the training phase results in optimal values for all free parameters in the CAE.}

\subsection{Variational autoencoders}

As opposed to CAEs, the variational autoencoder (VAE) \cite{kingma2013auto} takes a Bayesian approach to latent-space modeling. We utilize a convolutional VAE architecture, where outer convolutional, pooling and upscaling layers are identical to CAE. The difference only arises in the bottleneck. \textcolor{black}{In a CAE, the bottleneck has a unique representation (of latent-space dimension $d_l$) for every input datapoint, whereas in a VAE these representations will be probability distribution functions (PDFs). This latent-space distribution is often parametrized by, for instance, a Normal distribution $N(\mu, \sigma^2)$ in our implementation: hence, the encoder will be a map from input points to the means $\mu$ and standard deviations $\sigma$ corresponding to latent-space PDFs. The number of parameters at the bottleneck will be $2\times d_l$, double that of a corresponding CAE. Similarly, a CAE decoder would be a deterministic map from discrete latent-space variables to target images. Whereas in VAE, a sample will be drawn from the continuous latent-space distribution $N(\mu, \sigma^2)$ to reconstruct the image.}

\begin{figure}
    \centering{\includegraphics[width=\textwidth]{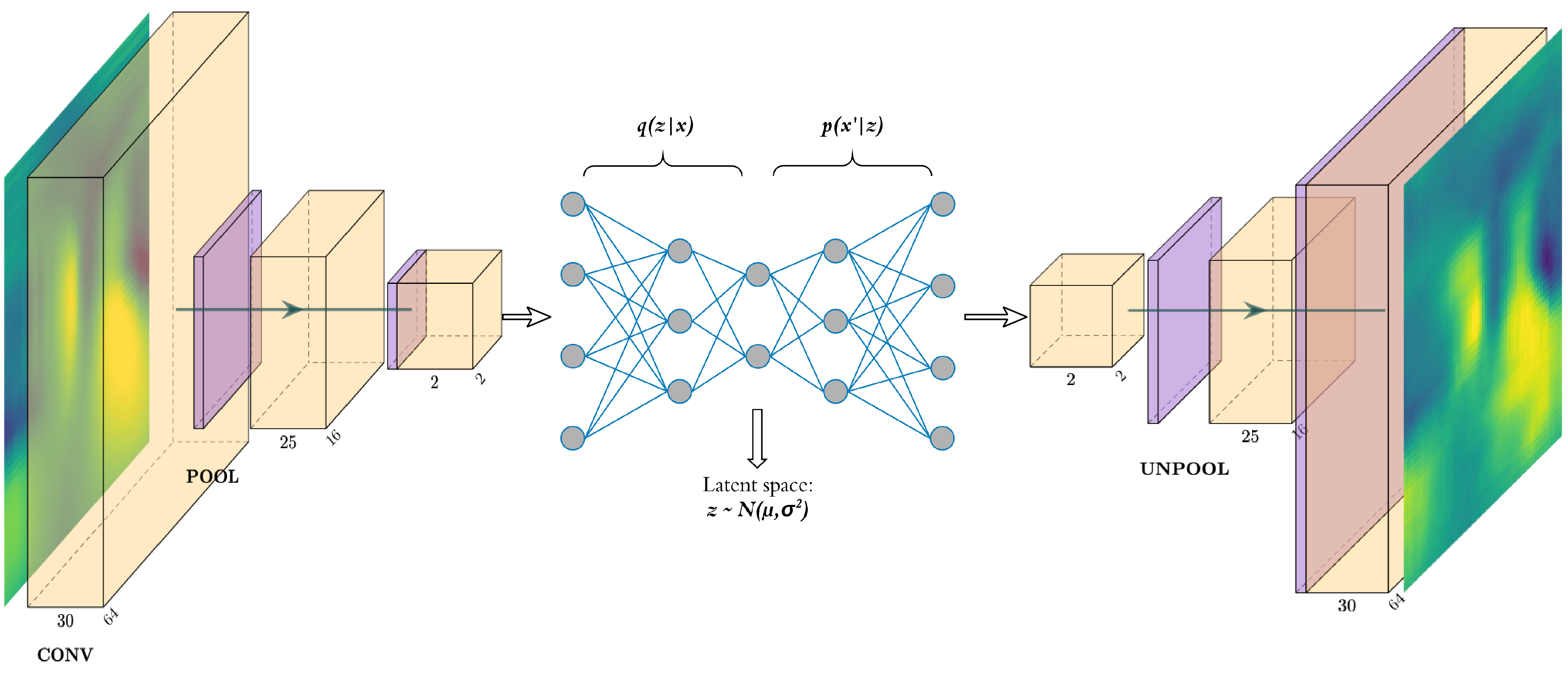}}
    \caption{Convolutional VAE architecture: Convolutional parts of the encoder and decoder are similar to that of a convolutional autoencoder. The bottleneck includes a mapping onto a normal distribution of latent-space variables. Sampling from this space is performed and decoded for generation of new data.}
    \label{vae_arch}
\end{figure}

\textcolor{black}{The final layers of a VAE encoder} are fully connected layers that project the input $\mathbf{x} \in \mathbb{R}^n $ onto the latent-variable $z$ space. Thus, the encoder effectively is a function $q(\mathbf{z}|\mathbf{x})$. The fully connected part of the decoder network  $p(\mathbf{x}^{\prime}|\mathbf{z})$ generates newly sampled $\mathbf{x}^{\prime}$ from the latent space. The output of this undergoes upsampling and convolutions (similar to that of the CAE) to reconstruct the image. The VAE also constrains the latent-space variables to follow a normal distribution $z \sim \mathcal{N}(\mathbf{\mu}, \mathbf{\sigma}^2)$ via the Kullback–Leibler divergence (KL divergence) term $D_{\mathrm{KL}}(q(\mathbf{z}|\mathbf{x_i})||p(\mathbf{z}|\mathbf{x}_i))$. The architecture of VAE (shown in Figure \ref{vae_arch}) is similar to that of the conventional CAE shown in Figure \ref{cae_arch}, except at the bottleneck, where the encoder network outputs the mean $\mu$ and variance $\sigma^2$.

The inference itself is undertaken using \emph{variational inference} (VI) by modeling the true distribution $q(\mathbf{z}|\mathbf{x})$ using a simple Gaussian distribution, and then minimizing the difference via the KL divergence as an addition to the loss function $E_{q(\mathbf{z}|\mathbf{x}_i)}[\log{p(\mathbf{z},\mathbf{x}_i)-\log{q(\mathbf{z}|\mathbf{x}_i)}}]$. The KL-divergence loss is applied such that the distribution on $z$ is as close to the normal distribution as possible.

With the inclusion of additional complexity to the loss function and the bottleneck, VAEs have more parameters to tune than CAEs. However, this also gives a significant control over the latent-space distribution. Depending on the data, dimensionality reduction using CAEs may not allow for a straightforward interpolation due to the presence of discontinuous clusters in the latent-space representation. On the other hand, VAEs constrain the latent-space representation to follow a pre-specified distribution, hence by design, facilitate easier interpolation. \textcolor{black}{In practice, the performance of CAEs and VAEs will depend on the quantity, quality, sampling and the behaviour of the training data. In applications with well-behaved data, CAEs may have the upper-hand in reconstruction quality due to their simpler network architecture. Simpler CAE architectures result in a latent-space representation that can be more easily interpolated upon, thus circumventing the need for an additional control the VAE provides at the expense of a more rigorous training. This in fact is the case with our applications shown in Section \ref{sec:experiments}.}

\section{Gaussian process regression}

Dimensionality reduction performed using POD, CAE or VAE techniques results in a representation of the original data, along with a model (the POD bases or decoders) to reconstruct the data for any point in the latent space. We then only require a temporal interpolation scheme fitted on the representation space, so that a time evolution of the dynamical system can be reconstructed, as schematically shown in Fig \ref{schematic}. In our  approach, we deploy the use of GPs \cite{williams2006gaussian} as our interpolation algorithm. While GPs perform Bayesian regression tasks with a high level of interpretabilty, they are generally computationally expensive for large data sizes. We achieve a considerable reduction of computational cost by fitting in the space of reduced dimensions. In addition, the GPR may also be restricted to a less noisy space compared with the original data set of the dynamical evolution.

A GP is an accumulation of random variables, every finite collection of which follows a multivariate Gaussian distribution. It can be perceived as a generalization of a multivariate Gaussian distribution with infinite space. GPs are a popular choice for non-linear regression \cite{williams1996gaussian} due to their flexibility and ease of use. In addition, one of their main advantages, is that they incorporate a principled way of measuring the uncertainty information, since they provide predictions in distributional form. For the purpose of this paper, a GPR model is used to fit the reduced space from the data-compression algorithms of Section~\ref{sec:data_compression}. Subsequently, the mean prediction, which corresponds to the maximum a posteriori (MAP) estimate, is used for the reconstructions. We use the GPflow library for the experiments \cite{de2017gpflow}.

A GP can be completely specified by its second-order statistics. Thus, a mean function $m(\mathbf{x})$ equal to zero can be assumed, and a positive definite covariance function (kernel) $k(\mathbf{x}, \mathbf{x'})$, which can be perceived as a measure of similarity between $\mathbf{x}$ and $\mathbf{x'}$, is the only requirement to specify the GP.

For a GPR model, we considered a GP $f$ and noisy training observations $\mathbf{y}$ of $n$ datapoints $\mathbf{x}$, derived from the true values $f(\mathbf{x})$ with additive i.i.d.~Gaussian noise $\epsilon$ with variance $\sigma_n^2$. In mathematical form, that is:
\begin{linenomath*}
\begin{align}
\begin{gathered}
\mathbf{y} = f(\mathbf{x}) + \epsilon, \\
\epsilon \sim \mathcal{N}(0, \sigma_n^2)\\
f(\mathbf{x}) \sim \operatorname{GP}(0, k(\mathbf{x},\mathbf{x'})),
\end{gathered}
\end{align}
\end{linenomath*}
where $k(\cdot,\cdot)$ is the kernel. \textcolor{black}{Effectively, it is assumed that for a GP $f$ observed at $\mathbf{x}$, the vector $f(\mathbf{x})$ is one sample from a multivariate Gaussian distribution of dimension equal to the number of datapoints $n$}. We obtain the complete specification of the GP, by maximizing the \textit{marginal likelihood}, which we can acquire by integrating the product of the Gaussian likelihood and the GP prior over $f$:
\begin{linenomath*}
\begin{align}
\begin{gathered}
p(\mathbf{y}|\mathbf{x}) = \int_f p(\mathbf{y}|f, \mathbf{x})p(f|\mathbf{x}) \,\mathrm{d}f.
\end{gathered}
\end{align}
\end{linenomath*}
For testing input $\mathbf{x_\star}$ and testing output $\mathbf{f_\star}$, we derive the joint marginal likelihood:
\[\begin{bmatrix}
\mathbf{y} \\
\mathbf{f_\star}
\end{bmatrix}\sim \mathcal{N}\left(\begin{bmatrix}
0 \\
0
\end{bmatrix},\begin{bmatrix}
k(\mathbf{x},\mathbf{x}) + \sigma_n^2 \mathbf{I} & k(\mathbf{x},\mathbf{x_\star}) \\
k(\mathbf{x_\star},\mathbf{x}) & k(\mathbf{x_\star},\mathbf{x_\star})
\end{bmatrix}\right),
\]
where $\mathbf{I}$ is the identity matrix.

Finally, by conditioning the joint distribution on the training data and the testing inputs, we derive the predictive distribution
\begin{linenomath*}
\begin{equation}
\mathbf{f_\star}|\mathbf{x}, \mathbf{x_\star}, \mathbf{y} \sim \mathcal{N} (\mathbf{\bar{f}}_\star, \operatorname{cov}(\mathbf{f_\star})),    
\end{equation}
\end{linenomath*}
where
\begin{linenomath*}
\begin{gather}
\begin{aligned}
\mathbf{\bar{f}}_\star &= k(\mathbf{x_\star},\mathbf{x})[k(\mathbf{x},\mathbf{x}) + \sigma_n^2 \mathbf{I}]^{-1}\mathbf{y} \\
\operatorname{cov}(\mathbf{f_\star}) &= k(\mathbf{x_\star},\mathbf{x_\star}) - k(\mathbf{x_\star},\mathbf{x})[k(\mathbf{x},\mathbf{x}) + \sigma_n^2 \mathbf{I}]^{-1}k(\mathbf{x},\mathbf{x_\star}).\label{predictive_dist}\\
\end{aligned}
\end{gather}
\end{linenomath*}

For the experiments in Section \ref{sec:experiments} we initially experimented with changepoint kernels \cite{lloyd2014automatic}. The intuition came from the data (see e.g. Figure~\ref{FR_Time}), where two typical behaviours are commonly observed; a steep increase or decrease of the latent dimension values for early times, and a subsequent, smoother change in direction, that eventually leads to stabilization. At first we examined a changepoint kernel only for the time feature, which was then added to a regular kernel that accounted for the other variables. The results were discouraging, which we attribute to the fact that this kernel structure leads to loss of correlational information between time and the other variables. Subsequently, we examined changepoint kernels that accounted for all parameters in both their sub-kernels. Even though this type of kernel was successful in producing acceptable results, and also detecting adequately the position of the changepoint, the output was only a slight improvement when compared to \textcolor{black}{a single Matérn 3/2 kernel}, which was our final choice. Furthermore, the computational time was substantially larger, which led us to abandon the pursuit of a changepoint kernel. \textcolor{black}{Ideally, due to the presence of two different time scales, an additive kernel with different scaling would intuitively seem better. However, the other GP interpolation dimensions are in the parameter space of the shallow water equations, which do not exhibit multiple length-scale dynamics. We attemped multiple configurations of additive kernels, including shared kernel length scales across parameter and time dimensions, individual kernel length scales (Automatic Relevance Detection (ARD)) for each dimension, and ARD only across the time dimension with shared kernel length scales for the parameter dimensions. However, none of these configurations were significantly different, in terms of results, from the single Matérn kernels. We suspect that additive kernels may required more training data as there are more parameters to fit. In the end we opted for the simplicity of the Matérn 3/2 kernel with length scale $l$, which gives good results with a lower number of training data points (desirable for ROM applications)}:
\begin{linenomath*}
\begin{align}
\begin{gathered}
k(\mathbf{x}, \mathbf{x'}) = \left(1 + \frac{\sqrt{3(\mathbf{x}-\mathbf{x'})^2}}{l}\right) \exp\left(-\frac{\sqrt{3(\mathbf{x}-\mathbf{x'})}}{l} \right).
\end{gathered}
\end{align}
\end{linenomath*}
We chose the Matérn 3/2 kernel due to its versatility, flexibility and smoothness, and more specifically its automatic relevance determination (ARD) extension \cite{bishop2006pattern}, which incorporates a separate parameter for each input variable and gave a significant improvement in our results.

\textcolor{black}{For the special case of a Matérn 3/2 kernel, Equation \eqref{predictive_dist} simplifies to:
\begin{linenomath*}
\begin{gather}
\begin{aligned}
\mathbf{\bar{f}}_\star &= \left(1 + \frac{\sqrt{3(\mathbf{x_\star}-\mathbf{x})^2}}{l}\right) \exp\left(-\frac{\sqrt{3(\mathbf{x_\star}-\mathbf{x})}}{l} \right)[ (1 + \sigma_n^2) \mathbf{I}]^{-1}\mathbf{y} \\
\operatorname{cov}(\mathbf{f_\star}) &= 1 - \left(1 + \frac{\sqrt{3(\mathbf{x_\star}-\mathbf{x})^2}}{l}\right) \exp\left(-\frac{\sqrt{3(\mathbf{x_\star}-\mathbf{x})}}{l} \right)[ (1+ \sigma_n^2) \mathbf{I}]^{-1}\left(1 + \frac{\sqrt{3(\mathbf{x}-\mathbf{x_\star})^2}}{l}\right) \exp\left(-\frac{\sqrt{3(\mathbf{x}-\mathbf{x_\star})}}{l} \right).\\
\end{aligned}
\end{gather}
\end{linenomath*}}

During the reconstruction phase, we focus on the predictions that correspond to $\mathbf{\bar{f}}_\star$. \textcolor{black}{For a more detailed explanation regarding selection of kernels the reader is referred to \cite{duvenaud2014automatic}}. 

\section{Experiments}\label{sec:experiments}

In this section, we outline several experiments designed to assess the various compression frameworks and how they interface with latent-space emulation using our aforementioned \textcolor{black}{Gaussian processes (GPs)}. A first series of assessments is solely targeted at assessing the fidelity of reconstruction (i.e., which framework offers the most efficient compression). Following this, we interface latent-space representations of our compressed fields with GP emulators to obtain low-dimensional surrogates with embedded uncertainty quantification. Finally, we outline the ability for the GP emulators to predict the dynamics' evolution at finer temporal resolutions.

For the purpose of training the compression frameworks and the GP emulators, we generate \num{100} forward-model solves which are obtained by a Latin hypercube sampling of different initial conditions $w$ between \numlist{-0.5;0.5} for each dimension $\bar{x}$ and $\bar{y}$. For each of these \num{100} simulations, \num{10} evenly-spaced snapshots in time are obtained to construct our total data set (i.e., we have \num{1000} flow-fields for $\rho \eta$ of resolution $64{\times}64$). We remind the reader that the equation-based simulations require the solution of a system of PDEs (for $\rho \eta$, $\rho \eta u$ and $\rho \eta v$), but our emulators will be built from $\rho \eta$ information alone. We split the \num{100} simulations into \num{80} for training, \num{10} for validation and \num{10} for testing. The validation data set is primarily utilized for early stopping criteria in the deep neural network autoencoders. To ensure no ``data leakage", we avoid using the testing data for anything other than \emph{a posteriori assessments}. We also note that no cross-validation is performed for the training and validation data sets due to the sufficient quality of the training (as observed through low training losses). Note that the training and validation data are combined into one data set for training GP emulators. All statistical and qualitative assessments in the following will be shown for the testing data alone.

\subsection{Reconstruction}

We begin by assessing the ability of our different compression frameworks, i.e.~the \textcolor{black}{proper orthogonal decomposition (POD), convolutional autoencoder (CAE), and variational autoencoder (VAE).} This comparison is obtained by training multiple encoders with varying degrees of freedom (DOF) in the latent space. The results of these experiments can be seen in Table \ref{Table_Compression}. Here, we have chosen \numlist{2; 4; 8; 16; 30; 40} DOF for all of our compression frameworks and have compared the fidelity of the reconstruction. We use metrics given by the coefficient of determination ($R^2$), mean squared error (MSE), and mean absolute error (MAE) to compare the true and reconstructed fields for testing data sets. Our analysis of the metrics indicate that the CAE is able to reach \textcolor{black}{better} reconstruction accuracy with \textcolor{black}{fewer degrees of freedom in latent space} than both the POD and VAE. Both the VAE and CAE are seen to possess an advantage \textcolor{black}{(in terms of efficient compression)} over the POD, due to their ability to find a non-linear low-dimensional manifold. POD, instead, obtains a linear affine subspace of the high-dimensional system. Interestingly, with increasing DOF in latent space (\num{\sim 40}) the POD method is seen to outperform the VAE. We also note that the CAE and VAE frameworks obtain their peak accuracy at around 8 DOF in the latent space and
\textcolor{black}{that the further addition of latent degrees of freedom leads to negligible changes in accuracy.} We remark that this aligns with the lack of any guarantees on \textcolor{black}{monotonic} convergence with increasing DOF for these nonlinear compression frameworks. POD, instead, is guaranteed to converge with increasing latent-space dimensions.

\sisetup{scientific-notation = true}
\begin{table}[h]
\caption{Latent-space compression and reconstruction metrics for testing data. The CAE is seen to obtain high accuracy with relatively few DOF in comparison to the POD and VAE.}
\centering
\begin{tabular}{|c|c|c|c|c|c|c|}
\hline
\multicolumn{7}{|c|}{Coefficient of determination}                          \\ \hline
Model/Latent DOF & $2$   & $4$            & $8$            & $16$           & 30              & $40$                            \\ \hline
        POD              & \num{0.10}     & \num{0.30}     & \num{0.55}     & \num{ 0.69}     & \num{0.82}      & \num{0.87}    \\ \hline
        CAE              & \num{0.37}     & \num{0.87}     & \num{0.91}     & \num{ 0.88}     & \num{0.91}      & \num{0.91}    \\ \hline
        VAE              & \num{0.35}     & \num{0.66}     & \num{0.86}     & \num{ 0.83}     & \num{0.82}      & \num{0.79}    \\ \hline
        \multicolumn{7}{|c|}{Mean squared error}                                                                                \\ \hline
        Model/Latent DOF & $2$            & $4$            & $8$            & $16$            & $30$            & $40$          \\ \hline
        POD              & \num{0.0025}   & \num{0.0021  } & \num{0.0014  } & \num{0.0010   } & \num{0.00063 }  & \num{0.00045} \\ \hline
        CAE              & \num{0.0017 }  & \num{0.00034 } & \num{0.00025 } & \num{0.00031  } & \num{0.00026 }  & \num{0.00025} \\ \hline
        VAE              & \num{0.0017  } & \num{0.00084 } & \num{0.00036 } & \num{0.00043  } & \num{0.00045 }  & \num{0.00052} \\ \hline
        \multicolumn{7}{|c|}{Mean absolute error}                                                                               \\ \hline
        Model/Latent DOF & $2$            & $4$            & $8$            & $16$            & $30$            & $40$          \\ \hline
        POD              & \num{0.029  }  & \num{0.027  }  & \num{0.021 }   & \num{0.017  }   & \num{0.013    } & \num{0.011  } \\ \hline
        CAE              & \num{0.021  }  & \num{0.0090 }  & \num{0.0075}   & \num{0.0083 }   & \num{0.0075   } & \num{0.0076 } \\ \hline
        VAE              & \num{0.023  }  & \num{0.015  }  & \num{0.0094}   & \num{0.010  }   & \num{0.010    } & \num{0.011  } \\ \hline
\end{tabular}
\label{Table_Compression}
\end{table}
\sisetup{scientific-notation = false}

\begin{figure}
    \centering
    \mbox{
    \subfigure[ \textcolor{black}{$\rho \eta$} reconstruction at $t=0.03$]{\label{4_DOF_Contours_sb1}\includegraphics[width=0.8\textwidth]{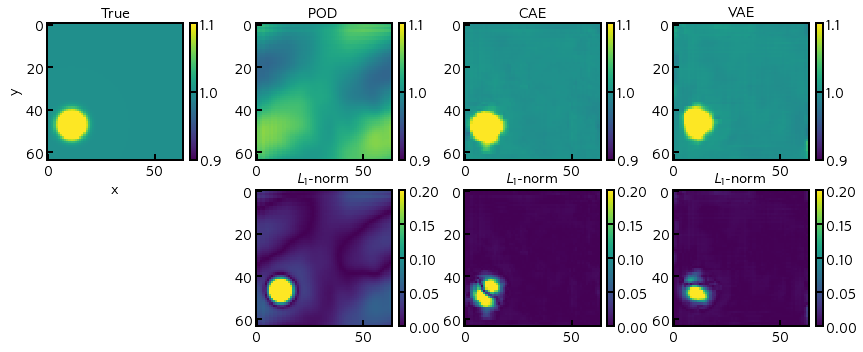}}
    }\\
    \mbox{
    \subfigure[\textcolor{black}{$\rho \eta$} reconstruction at $t=0.06$]{\label{4_DOF_Contours_sb2} \includegraphics[width=0.8\textwidth]{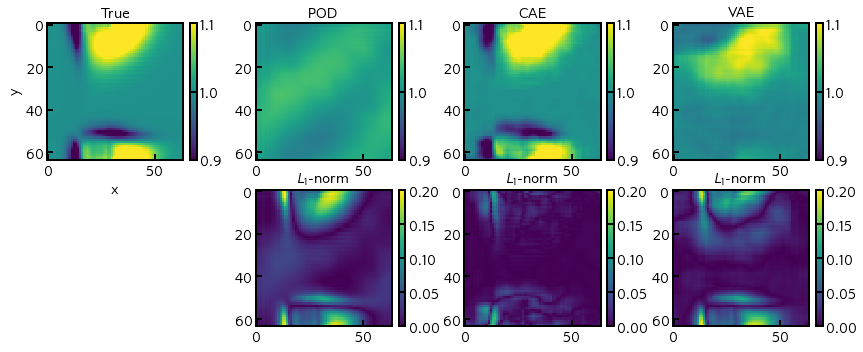}}
    }\\
    \mbox{
    \subfigure[\textcolor{black}{$\rho \eta$} reconstruction at $t=0.09$]{\label{4_DOF_Contours_sb3} \includegraphics[width=0.8\textwidth]{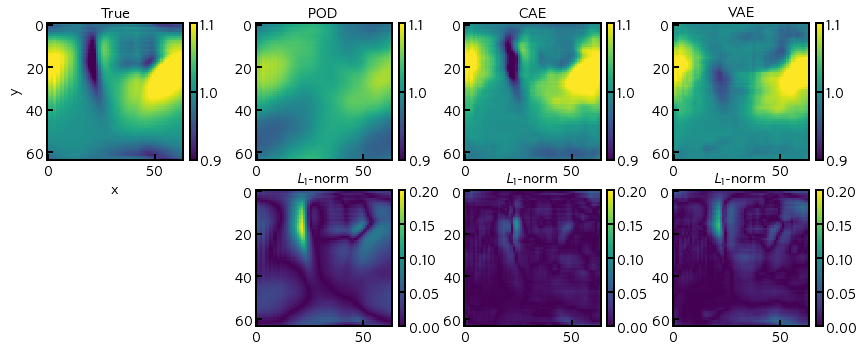}}
    }\\
    \caption{\textcolor{black}{$\rho \eta$} reconstructions \textcolor{black}{and $L_1$ errors} from true latent-space evolutions for a four-dimensional latent space for time $t=0.03$ (top), $t=0.06$ (middle), $t=0.09$ (bottom) for a testing initial condition. Qualitatively, for the same data set and similar architectures, the CAE outperforms the VAE. Note that both the VAE and CAE are superior to the POD.}
    \label{4_DOF_Contours}
\end{figure}

\begin{figure}
    \centering
    \mbox{
    \subfigure[\textcolor{black}{$\rho \eta$} reconstruction at $t=0.03$]{\label{8_DOF_Contours_sb1}\includegraphics[width=0.8\textwidth]{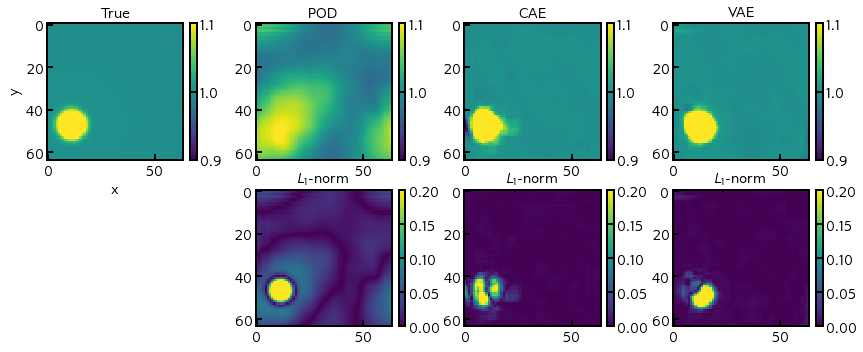}}
    }\\
    \mbox{
    \subfigure[\textcolor{black}{$\rho \eta$} reconstruction at $t=0.06$]{\label{8_DOF_Contours_sb2}\includegraphics[width=0.8\textwidth]{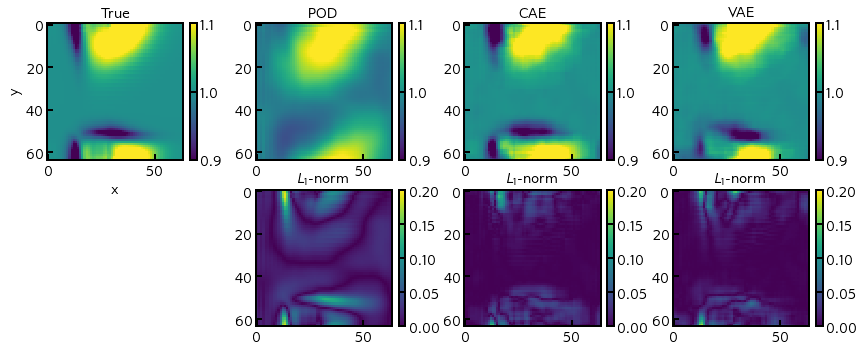}}
    }\\
    \mbox{
    \subfigure[\textcolor{black}{$\rho \eta$} reconstruction at $t=0.09$]{\label{8_DOF_Contours_sb3}\includegraphics[width=0.8\textwidth]{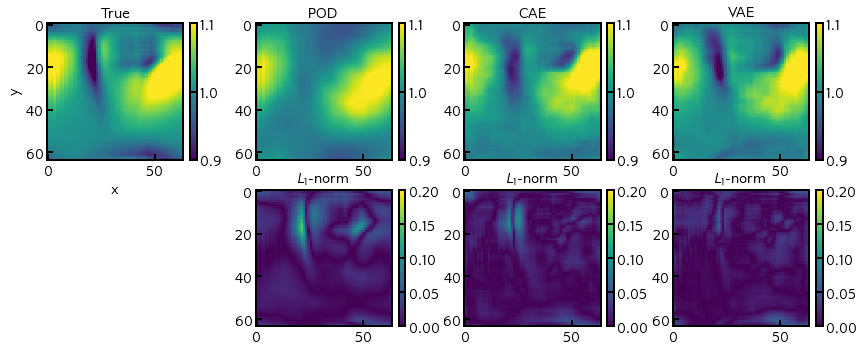}}
    }\\
    \caption{\textcolor{black}{$\rho \eta$} reconstructions \textcolor{black}{and $L_1$ errors} from true latent-space evolutions for an eight-dimensional latent space for time $t=0.03$ (top), $t=0.06$ (middle), $t=0.09$ (bottom) for a testing initial condition. Qualitatively, for the same data set and similar architectures, the CAE outperforms the VAE. Note that both VAE and CAE are superior to the POD.}
    \label{8_DOF_Contours}
\end{figure}

\begin{figure}
    \centering
    \mbox{
    \subfigure[\textcolor{black}{$\rho \eta$} reconstruction at $t=0.03$]{\includegraphics[width=0.8\textwidth]{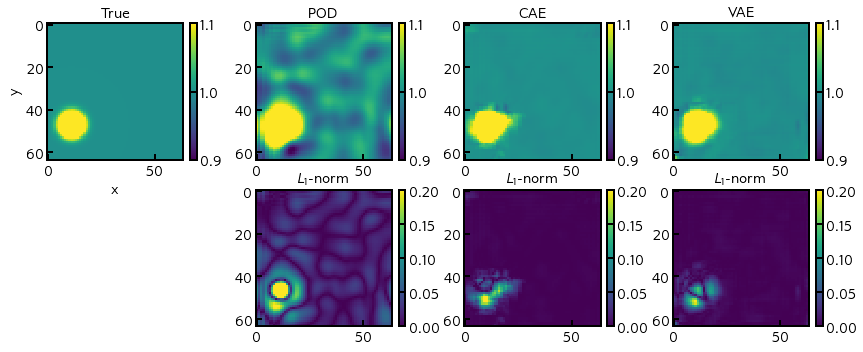}}
    }\\
    \mbox{
    \subfigure[\textcolor{black}{$\rho \eta$} reconstruction at $t=0.06$]{\includegraphics[width=0.8\textwidth]{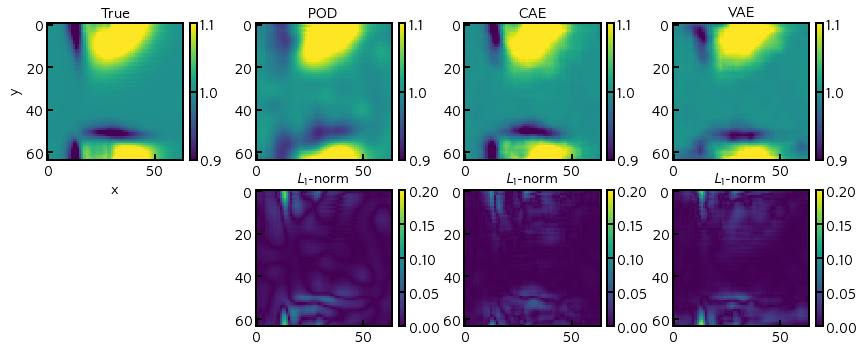}}
    }\\
    \mbox{
    \subfigure[\textcolor{black}{$\rho \eta$} reconstruction at $t=0.09$]{\includegraphics[width=0.8\textwidth]{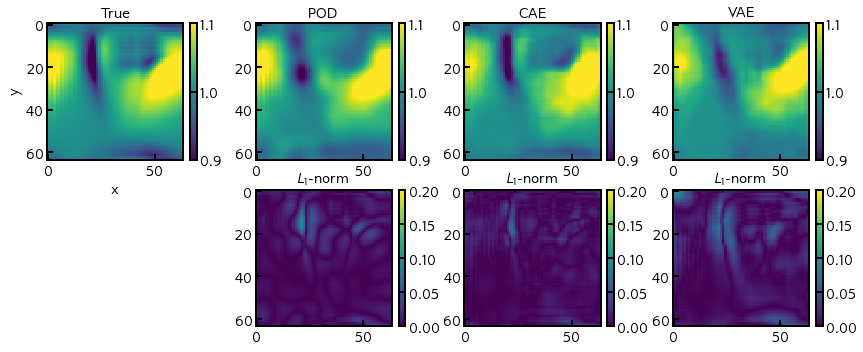}}
    }\\
    \caption{\textcolor{black}{$\rho \eta$} reconstructions \textcolor{black}{and $L_1$ errors} from true latent-space evolutions for a forty-dimensional latent space for time $t=0.03$ (top), $t=0.06$ (middle), $t=0.09$ (bottom) for a testing initial condition. Qualitatively, for the same data set and similar architectures, the CAE outperforms the VAE and is comparable to the POD encoding. This is because of the greater DOF in the latent space.}
    \label{40_DOF_Contours}
\end{figure}

Following these quantitative assessments, we assess the reconstruction fidelity of the different frameworks by comparing contours from the different methods with varying DOF in the latent space. Figure \ref{4_DOF_Contours} shows the performance for our three compression frameworks for four DOF in the latent space. At this coarse resolution (in the latent space), the linear compression of POD is inadequate at capturing the coherent features in the solution field upon reconstruction. This is due to the advective nature of the dynamics of this data set and the associated high Kolmogorov width. In contrast, both CAE and VAE are able to identify coherent structures in the flow field after being reconstructed from a latent space. \textcolor{black}{The reconstruction framework is successfully able to reconstruct the initial stages of the Gaussian pulse in Figure \ref{4_DOF_Contours_sb1} at time $t=0.03$ where the errors are mainly concentrated in the region of the pulse. Subsequent evolution of the flow shown at times $t=0.06$ and $t=0.09$ (shown in Figures \ref{4_DOF_Contours_sb2} and \ref{4_DOF_Contours_sb2}) show that errors are concentrated on the propagating perturbations.} Note, however, that the CAE is able to identify the crests and troughs in the flow field in a more accurate manner when compared with the POD and VAE. We observe similar results from an eight-dimensional latent space where the CAE and VAE are still seen to outperform POD \textcolor{black}{for times $t=0.03$, $t=0.06$, and $t=0.09$ in Figures \ref{8_DOF_Contours_sb1}, \ref{8_DOF_Contours_sb2}, and \ref{8_DOF_Contours_sb3} respectively}. For both cases (i.e., four and eight DOF), the CAE and VAE are seen to struggle with reconstructing the dissipating coherent features later in the evolution of the system. For completeness, we also show a result for a forty-dimensional latent space at different times in Figure \ref{40_DOF_Contours}, where POD can be seen to capture the spatio-temporal trends of the true solution in a much better fashion. Note that improvements in the CAE and VAE are marginal with the POD outperforming both these frameworks later in the evolution of system (at $t=0.09$).

\textcolor{black}{Finally, we assess the computational costs associated with training the different compression frameworks as shown in Table \ref{Training_times}. Observe that the computational cost of the POD-based compression is invariant to the magnitude of truncation. The autoencoder architectures are seen to be a factor of $5$ to $8$ times more expensive.}

\sisetup{scientific-notation = true}
\begin{table}[h]
\caption{Computational costs (in seconds) associated with training different compression frameworks. All assessments were performed on a single core of an Intel Core-i7 processor with a clock speed of \SI{1.90}{GHz}.}
\centering
\begin{tabular}{|c|c|c|c|c|c|c|}
\hline
\multicolumn{7}{|c|}{Training duration for compression framework (seconds)}                          \\ \hline
Model/Latent DOF & $2$   & $4$            & $8$            & $16$           & 30              & $40$                            \\ \hline
        POD              & \num{53}     & \num{53}     & \num{53}     & \num{ 53}     & \num{53}      & \num{53}    \\ \hline
        CAE              & \num{251}     & \num{478}     & \num{409}     & \num{359}     & \num{294}      & \num{264}    \\ \hline
        VAE              & \num{283}     & \num{637}     & \num{309}     & \num{172}     & \num{175}      & \num{346}    \\ \hline
\end{tabular}
\label{Training_times}
\end{table}
\sisetup{scientific-notation = false}

\subsection{Latent-space forecasts}

Next we test the ability of our trained GP emulators to forecast the evolution of systems in their latent-space representations. For this assessment, we choose trained encoders (with $4$, $8$ and $40$ latent-space dimensions) to obtain training and validation data for fitting our previously introduced GPs. Once trained, the GPs are tasked with predicting the evolution of the latent state \emph{in reduced space} for a set of test initial conditions.

\begin{figure}
    \centering
    \mbox{
    \subfigure[POD]{\includegraphics[width=0.9\textwidth]{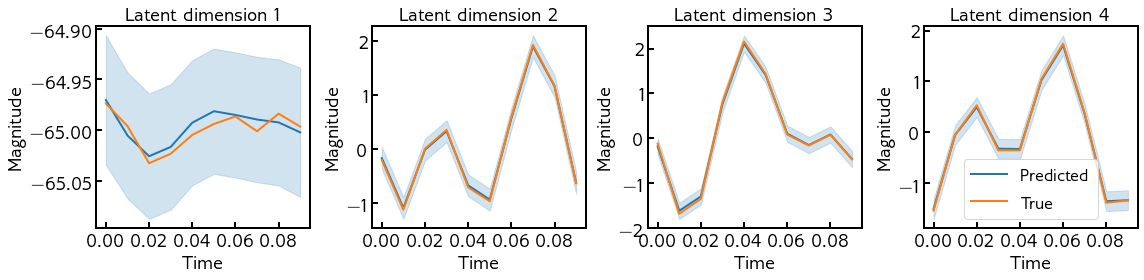}}
    } \\
    \mbox{
    \subfigure[CAE]{\includegraphics[width=0.9\textwidth]{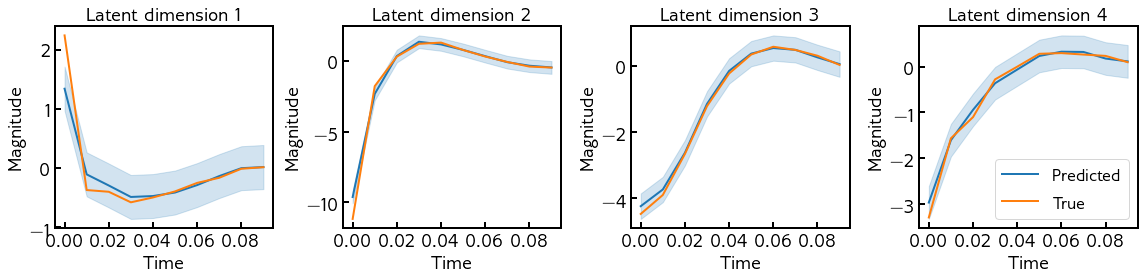}}
    } \\
    \mbox{
    \subfigure[VAE]{\includegraphics[width=0.9\textwidth]{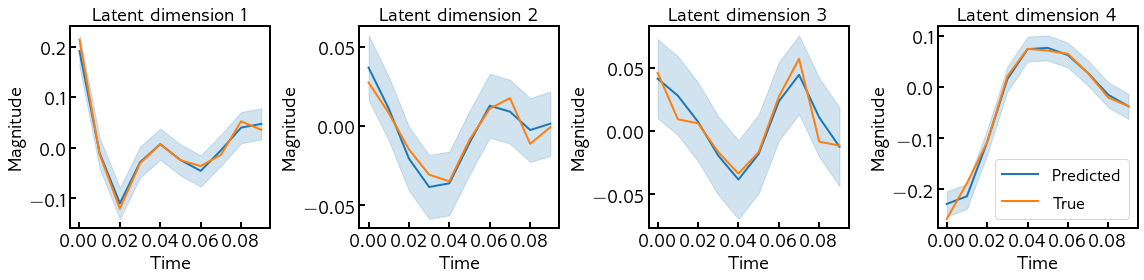}}
    } \\
    \mbox{
    \subfigure[Galerkin Projection]{\includegraphics[width=0.9\textwidth]{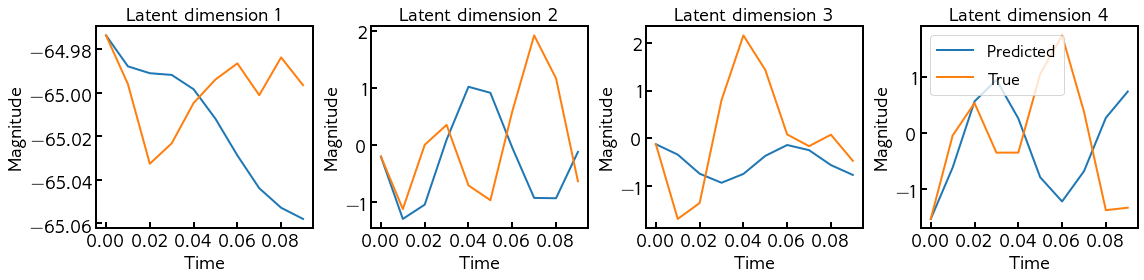}}
    }
    \caption{Latent-space forecasts on testing data for (a) POD, (b) CAE, (c) VAE, and (d) \textcolor{black}{the POD Galerkin Projection method} with a reduced dimension of 4 DOF. The shaded areas indicate confidence intervals from the probabilistic forecasts. \textcolor{black}{Note how the Galerkin projection precludes probabilistic estimates of latent-space evolution.}}
    \label{4_DOF_GP}
\end{figure}

\begin{figure}
    \centering
    \mbox{
    \subfigure[POD]{\includegraphics[width=0.9\textwidth]{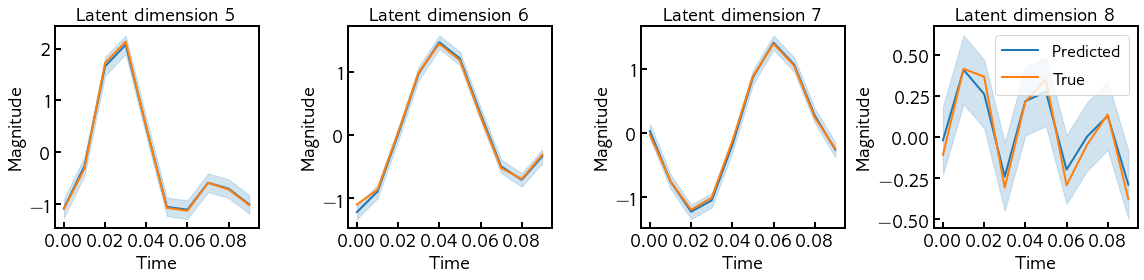}}
    } \\
    \mbox{
    \subfigure[CAE]{\includegraphics[width=0.9\textwidth]{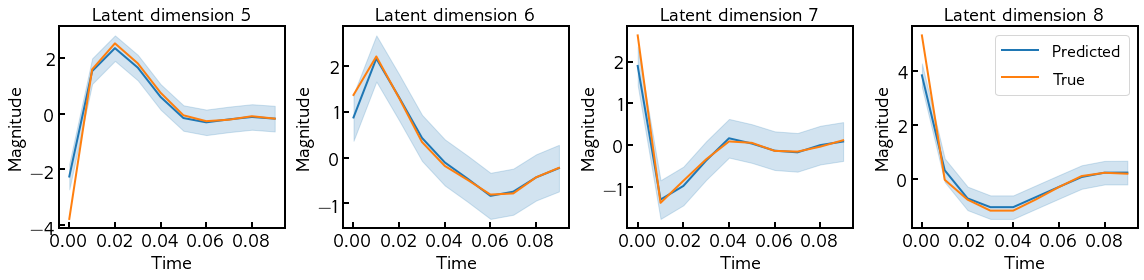}}
    } \\
    \mbox{
    \subfigure[VAE]{\includegraphics[width=0.9\textwidth]{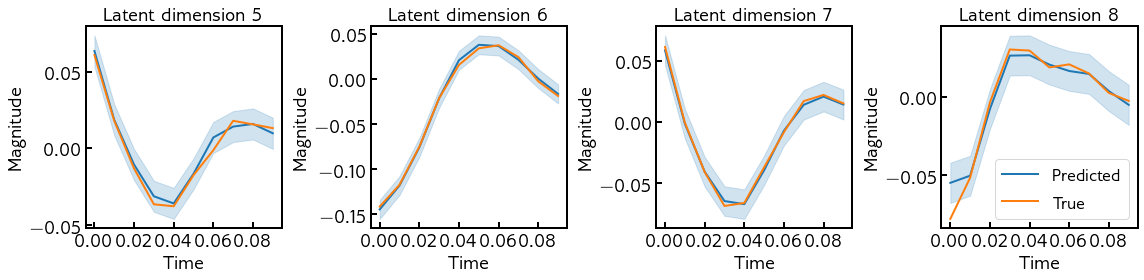}}
    } \\
    \mbox{
    \subfigure[Galerkin Projection]{\includegraphics[width=0.9\textwidth]{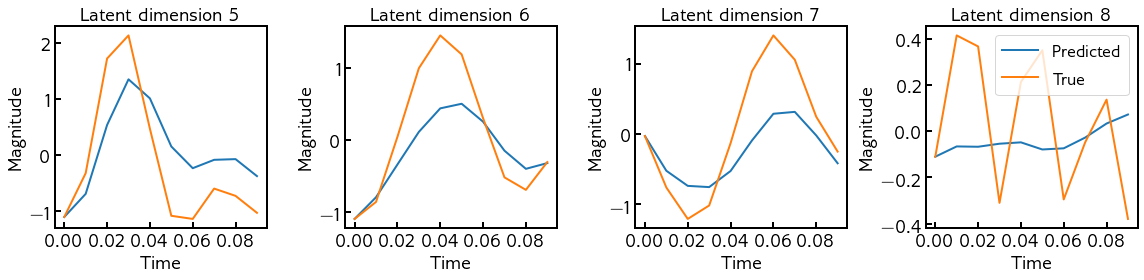}}
    }
    \caption{Latent-space forecasts on testing data for (a) POD, (b) CAE and (c) VAE, and (d) \textcolor{black}{the POD Galerkin Projection method} with a reduced dimension of 8 DOF. The shaded areas indicate confidence intervals from the probabilistic forecasts. \textcolor{black}{Note that we have shown the evolution of only 4 latent-space dimensions and that the Galerkin projection precludes probabilistic estimates of latent-space evolution.}}
    \label{8_DOF_GP}
\end{figure}

Figure \ref{4_DOF_GP} shows the latent-space evolution of a testing simulation over time for the three different compression methodologies. This result utilizes solely four latent dimensions. It is readily apparent that CAE compression leads to a smooth evolution of the system in the latent space. \textcolor{black}{This reflects the smooth behavior of the data that can be captured efficiently with the unconstrained latent space of the CAE}. In contrast, POD and VAE methods display significant oscillations. \textcolor{black}{The VAE is outperformed in this specific example, likely because four variables drawn from latent-space distributions are not sufficient to capture the time evolution in a smooth manner. Regardless,} the GP is able to capture the behavior of the system evolution and also provides confidence intervals which are based on two standard deviations around the mean. \textcolor{black}{The difference in the smoothness of the latent-space time evolution results different GP kernels and hyperparameters being optimized during latent-space fitting.}  With the exception of a few instances in time, the confidence intervals are able to envelope the true evolution of the system. We see similar results in Figure \ref{8_DOF_GP} where eight latent-space DOF are obtained for each of our compression frameworks. While the POD is seen to provide oscillatory system evolution (as before), the CAE and VAE are smooth. \textcolor{black}{Smoothness in the VAE latent space has improved, along with the reconstruction accuracy, showing that a lower compression (i.e., an increase of DOF from 4 to 8) has enabled better expressiveness. However, this trend may not necessarily be extrapolated to any higher DOFs, as over-fitting of noisy features in the data may occur.} In either case, the constructed GP is able to recover the evolution well. \textcolor{black}{For the purpose of comparison, we also provide reduced-order model predictions from the well-known POD Galerkin Projection methodology in both figures. The Galerkin projection is a standard model order reduction technique which projects the governing equations to the reduced orthonormal basis obtained by truncating the POD modes. Subsequently, one is left with a system of equations which are computationally inexpensive to solve in this basis. Further details of the Galerkin projection may be obtained at \cite{rowley2004model,san2015principal}. A key limitation of the Galerkin projection is the requirement of observations from all the conserved variables (i.e., $\rho \eta, \rho \eta u, \rho \eta v$) and the closure errors due to inaccurate capture of the dynamics of the finer frequencies (discarded by truncation of the POD modes for advection-dominated data sets). Another point to note is that the Galerkin projection is deterministic in nature and cannot provide probabilistic estimates which are useful in the presence of noisy data snapshots. Results from the $4$ and $8$ degree of freedom evolution of our system indicate the limitations of this approach for advection dominated dynamics.}

\textcolor{black}{We also compare the computational costs for training the GPs in the latent space for varying compression as shown in Table \ref{fidelities_cost}. As expected, the time to solution for a trained GP scales with the number of degrees of freedom in the latent space. Training the POD-based compressed representations is seen to be generally more expensive as more degrees of freedom are retained in the latent space. We hypothesize this to be caused by the greater fluctuations in the time series obtained by POD truncated representations of the flow field.}

\sisetup{scientific-notation = true}
\begin{table}[h]
\caption{Computational costs (in seconds) associated with training different GP interpolation frameworks. The POD latent-space training did not converge for this assessment. All assessments were performed on a single node of an Intel Core-I7 processor with a clockspeed of 1.90 GHz.}
\centering
\begin{tabular}{|c|c|c|c|c|c|c|}
\hline
Model/Latent DOF & $2$   & $4$  & $8$  & $16$  & $30$  & $40$                            \\ \hline
        POD              & \num{15}     & \num{29}     & \num{68}     & \num{156}     & \num{291}      & N/A   \\ \hline
        CAE              & \num{8}     & \num{32}     & \num{53}     & \num{101}     & \num{200}      & \num{291}    \\ \hline
        VAE              & \num{9}     & \num{22}     & \num{48}     & \num{104}     & \num{225}      & \num{253}    \\ \hline
\end{tabular}
\label{fidelities_cost}
\end{table}
\sisetup{scientific-notation = false}

\subsection{Reconstruction from latent-space forecasts}

\begin{figure}
    \centering
    \mbox{
    \subfigure[GP Reconstructions \textcolor{black}{for $\rho \eta$} at $t=0.03$]{\label{4_DOF_t1_GP_sb1}\includegraphics[width=0.9\textwidth]{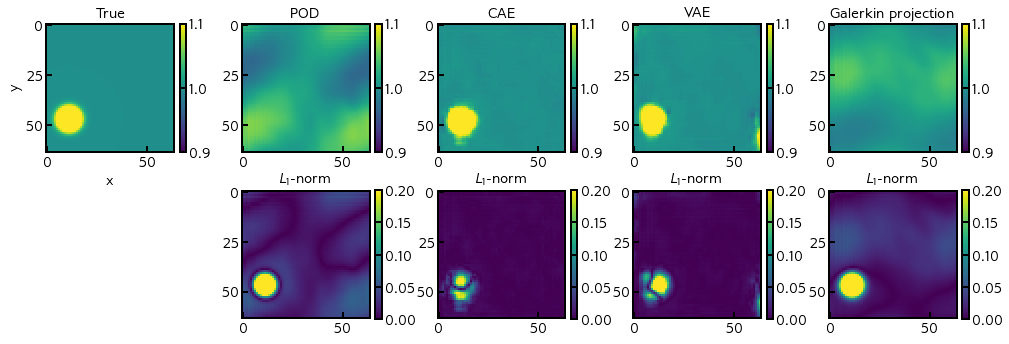}}
    }\\
    \mbox{
    \subfigure[GP Reconstructions \textcolor{black}{for $\rho \eta$} at $t=0.06$]{\label{4_DOF_t1_GP_sb2}\includegraphics[width=0.9\textwidth]{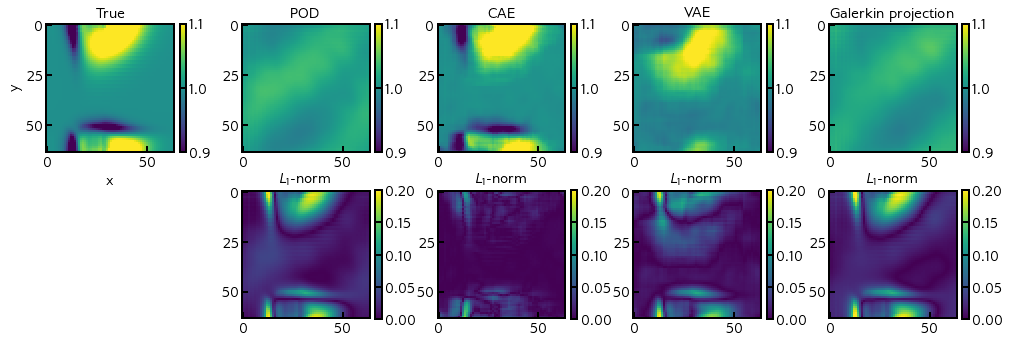}}
    }\\
    \mbox{
    \subfigure[GP Reconstructions \textcolor{black}{for $\rho \eta$} at $t=0.09$]{\label{4_DOF_t1_GP_sb3}\includegraphics[width=0.9\textwidth]{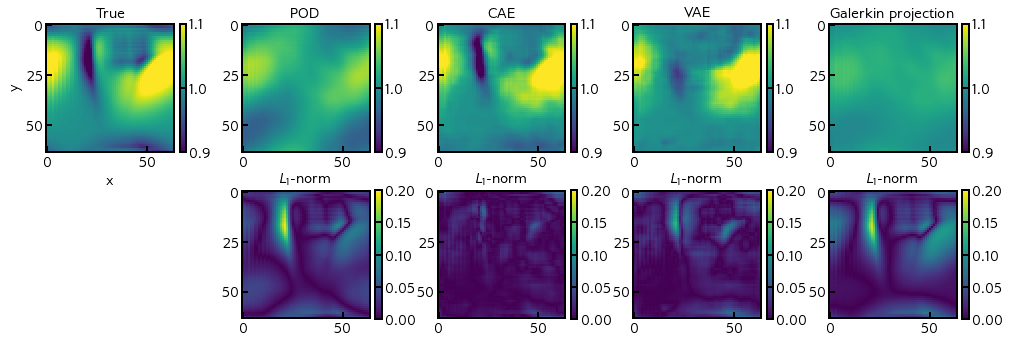}}
    }\\
    \caption{\textcolor{black}{$\rho \eta$} reconstructions \textcolor{black}{and $L_1$ errors} from GP forecasts for a four-dimensional latent space for time (a) $t=0.03$, (b) $t=0.06$ and (c) $t=0.09$ for a testing initial condition. Qualitatively, for the same data set and similar architectures, the CAE outperforms the VAE. Note that both the VAE and CAE are superior to the POD.}
    \label{4_DOF_Contours_GP}
\end{figure}

\begin{figure}
    \centering
    \mbox{
    \subfigure[GP Reconstructions \textcolor{black}{for $\rho \eta$} at $t=0.03$]{\label{8_DOF_t1_GP_sb1}\includegraphics[width=0.9\textwidth]{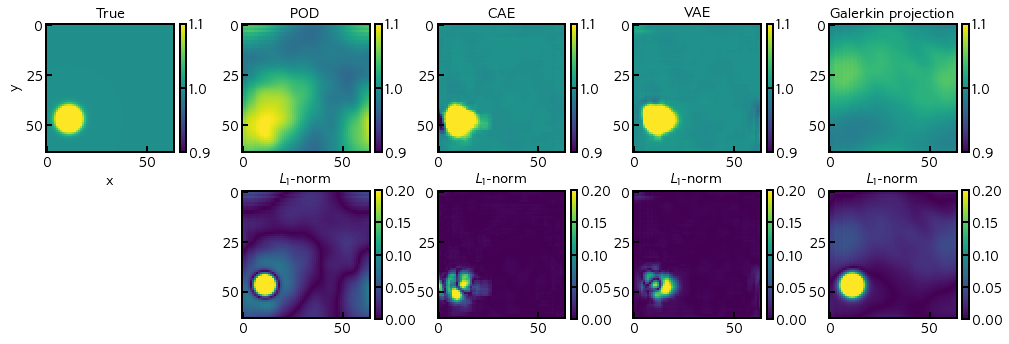}}
    }\\
    \mbox{
    \subfigure[GP Reconstructions \textcolor{black}{for $\rho \eta$} at $t=0.06$]{\label{8_DOF_t1_GP_sb2}\includegraphics[width=0.9\textwidth]{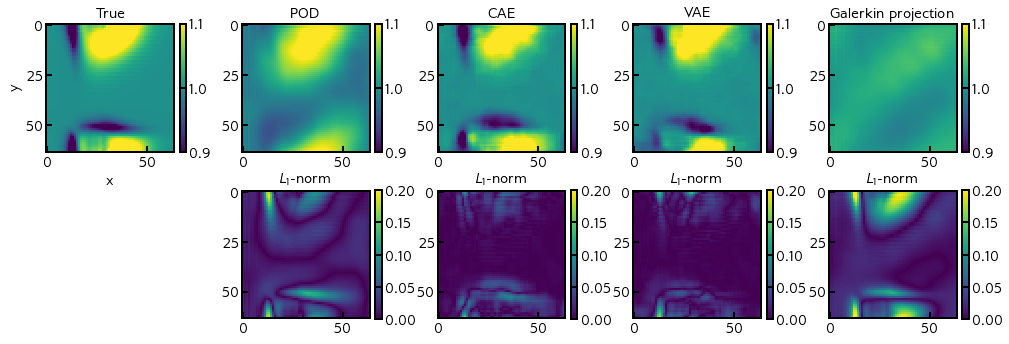}}
    }\\
    \mbox{
    \subfigure[GP Reconstructions \textcolor{black}{for $\rho \eta$} at $t=0.09$]{\label{8_DOF_t1_GP_sb3}\includegraphics[width=0.9\textwidth]{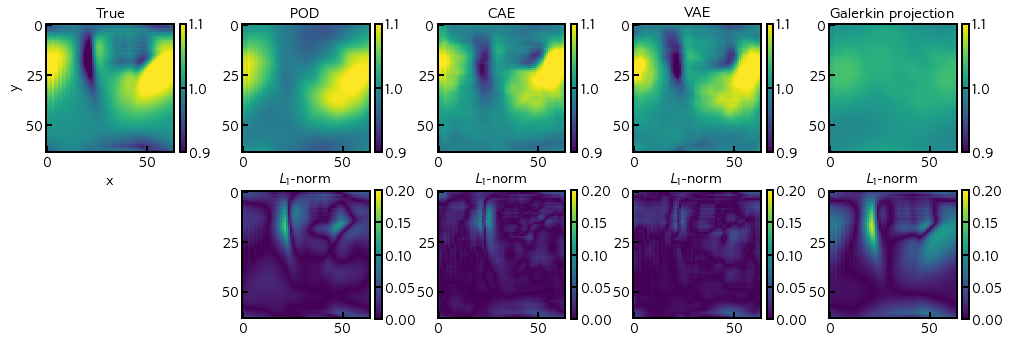}}
    }\\
    \caption{\textcolor{black}{$\rho \eta$} reconstructions \textcolor{black}{and $L_1$ errors} from GP forecasts for an eight-dimensional latent space for time (a) $t=0.03$, (b) $t=0.06$ and (c) $t=0.09$ for a testing initial condition. Qualitatively, for the same data set and similar architectures, the CAE outperforms the VAE. Note that both VAE and CAE are superior to the POD.}
    \label{8_DOF_Contours_GP}
\end{figure}

We proceed by assessing the fidelity of the reconstructions from the latent space using the GP forecasts of the previous subsection. Qualitative comparisons for the reconstruction of a test simulation are shown in \textcolor{black}{Figure \ref{4_DOF_Contours_GP} for a four-dimensional latent space at three different times $t=0.03$ (Figure \ref{4_DOF_t1_GP_sb1}), $t=0.06$ (Figure \ref{4_DOF_t1_GP_sb2}), and $t=0.09$ (Figure \ref{4_DOF_t1_GP_sb3})}. The CAE and VAE compression is seen to outperform the linear reduced-basis constructed by POD. This aligns with past studies where nonlinear compression methods have outperformed the POD. The CAE is seen to be more accurate than the VAE due to its deterministic formulation during compression. We observe the CAE and VAE to outperform the POD for the eight-dimensional latent space as well, as shown for different times $t=0.03$ (Figure \ref{8_DOF_t1_GP_sb1}), $t=0.06$ (Figure \ref{8_DOF_t1_GP_sb2}), and $t=0.09$ (Figure \ref{8_DOF_t1_GP_sb3}). Qualitatively, the \textcolor{black}{use of a CAE-based GP interpolation} is seen to be the best compression technique of all the methods \textcolor{black}{particularly for instances where there are greater number of latent dimensions}. Table \ref{Table_Metrics_Test} shows different metrics to establish these conclusions quantitatively. Note that all these metrics are evaluated on the reconstructed data in physical space.

\sisetup{scientific-notation = true}
\begin{table}[ht!]
    \centering
    \caption{Metrics obtained via GP-based forecast in the latent space followed by reconstruction for testing data.}
    \begin{tabular}{|c|c|c|c|c|c|}
        \hline
        \multicolumn{6}{|c|}{Coefficient of determination} \\ \hline
        Model/DOF & $4$             & $8$ & $16$ & $30$ & $40$                 \\ \hline
        POD       & \num{-2.14}    & \num{-0.17} & \num{0.29} & \num{0.54} & N/A        \\ \hline
        CAE       & \num{0.834}    & \num{0.786} & \num{0.857} & \num{0.864} & \num{0.854} \\ \hline
        VAE       & \num{0.558}    & \num{0.804} & \num{0.646} & \num{0.798} & \num{0.754} \\ \hline
        Galerkin Projection   & \num{-11.44} & \num{-3.23} & \num{-1.25} & \num{-0.348} & \num{0.109}    \\ \hline
        \multicolumn{6}{|c|}{Mean squared error}           \\ \hline
        Model/DOF & $4$             & $8$  & $16$ & $30$ & $40$                \\ \hline
        POD       & \num{0.028   }  & \num{0.022} & \num{0.019} & \num{0.016} & N/A  \\ \hline
        CAE       & \num{0.0099 }  & \num{0.011} & \num{0.0086} & \num{0.0082} & \num{0.0088} \\ \hline
        VAE       & \num{0.016 }  & \num{0.010}  & \num{0.014} & \num{0.011} & \num{0.013} \\ \hline
        Galerkin Projection   & \num{0.036}  & \num{0.027} & \num{0.025} & \num{0.022} &  \num{0.019}  \\ \hline
        \multicolumn{6}{|c|}{Mean absolute error}          \\ \hline
        Model/DOF & $4$             & $8$ & $16$ & $30$ & $40$                 \\ \hline
        POD       & \num{0.0022    } & \num{0.0016} & \num{0.0012} & \num{0.00092} & N/A    \\ \hline
        CAE       & \num{0.00043   } & \num{0.00044} & \num{0.00033} & \num{0.00032} & \num{0.00032}    \\ \hline
        VAE       & \num{0.00081    } & \num{0.00047} & \num{0.00072} & \num{0.00041} & \num{0.00058} \\ \hline
        Galerkin Projection   & \num{0.0031} & \num{0.0020} & \num{0.0017} & \num{0.0014} &  \num{0.0011} \\ \hline
    \end{tabular}
    \label{Table_Metrics_Test}
\end{table}
\sisetup{scientific-notation = false}

\textcolor{black}{The shallow water equation system is used as a model problem for the demonstration of our surrogate modeling frameworks. In particular, it is important to demonstrate the potential benefits during online deployment of these surrogate modeling mechanisms when the degrees of freedom of the discretized PDEs become extremely large. Table \ref{FOM_Times} shows how the utilization of finer discretizations leads to unfavorable compute costs. The proposed surrogate modeling strategies may then be used for efficient cost amortization by moving online costs for generating training data and utilized the reduced-order model for tasks such as control and optimization.}

\sisetup{scientific-notation = true}
\begin{table}[h]
\caption{Online computational costs (in seconds) for deployment of reduced-order framework full-order model (FOM) at varying fidelities. All assessments were performed on a single core of an Intel Core-i7 processor with a clockspeed of \SI{1.90}{GHz}.}
\centering
\begin{tabular}{|c|c|c|c|c|c|}
\hline
\multicolumn{6}{|c|}{Computational costs for 8-DOF ROM and various grid resolutions. }  \\ \hline
GP: 8 DOF & FOM: $32\times32$ & FOM: $64\times64$  & FOM: $128\times128$  & FOM: $256\times256$  & FOM: $512\times512$ \\ \hline
        \num{0.46} & \num{0.6}     & \num{1.6}     & \num{7.1}     & \num{150}     & \num{798} \\ \hline
\end{tabular}
\label{FOM_Times}
\end{table}
\sisetup{scientific-notation = false}

\subsection{Temporal super-resolution}

It must be highlighted that ML-based time-series forecasting methods are generally formulated in a discrete fashion where the temporal resolution of the training data determines the resolution of the ROM deployment. Also, most studies of ML-based forecasting in time assume a regular sampling of the state in time. In practice, due to simulation or experimental limitations, state information may be available sparsely in time and at irregular intervals. Therefore, the construction of a ROM that is continuous in the temporal variable allows us to sample at intermediate time steps with quantified uncertainty. Therefore, we test the ability of our parameterized non-intrusive ROM for interpolation \emph{in time}. In this assessment, we establish the performance of the ROM to sample at locations (in time) that were not obtained in the discrete training and testing data. This is made possible due to the continuous function approximation property of the GPR. 

\begin{figure}
    \centering
    \includegraphics[width=\textwidth]{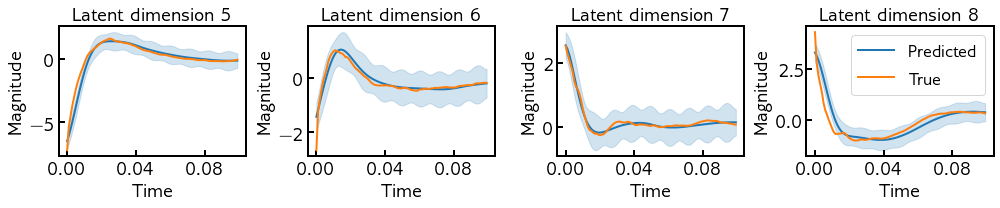}
    \caption{Latent-space interpolation for a testing data set with finer resolution. \textcolor{black}{Here we show latent dimensions 5-8 for an 8-dimensional latent space}. The purpose of this assessment is to determine the proposed framework's viability for super-resolution in time. While all the training data is obtained by sampling a simulation only \num{10} times over the course of its evolution, the assessments are tested against a simulation that has been sampled \num{100} times.}
    \label{FR_Time}
\end{figure}
\begin{figure}
    \centering
    \mbox{
    \subfigure[\textcolor{black}{$\rho \eta$} reconstruction at $t=0.005$]{\includegraphics[width=\textwidth]{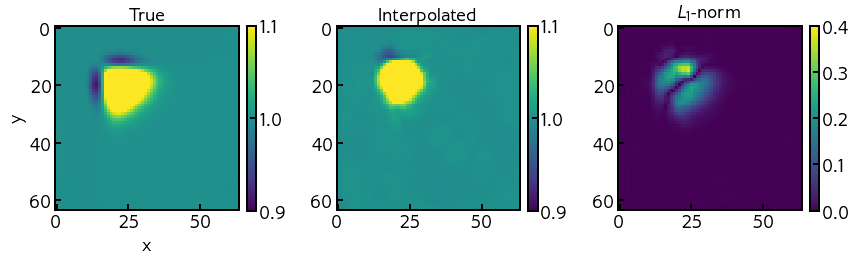}}
    }\\
    \mbox{
    \subfigure[\textcolor{black}{$\rho \eta$} reconstruction at $t=0.035$]{\includegraphics[width=\textwidth]{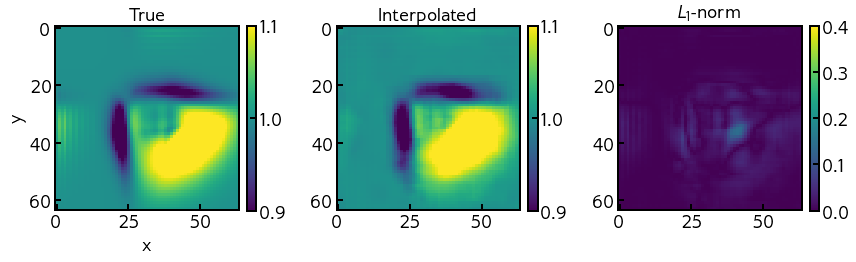}}
    }\\
    \mbox{
    \subfigure[\textcolor{black}{$\rho \eta$} reconstruction at $t=0.065$]{\includegraphics[width=\textwidth]{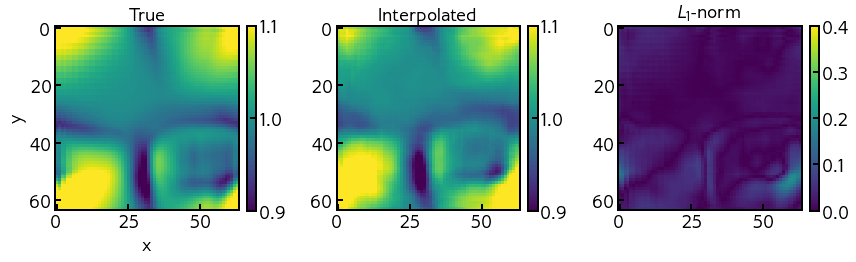}}
    }\\
    \caption{True fields (left), temporally interpolated predictions (middle) and $L_1$-norm errors with truth (right) for a testing \textcolor{black}{$\rho \eta$} that has been sampled \num{10} times more finely. Note that the shown instances in time are not the usual sampling points utilized for our CAE or GP training.}
    \label{FR_Viz}
\end{figure}

To assess this capability, we re-generate our testing data which is now sampled \num{10} times more finely in the temporal dimension. We utilize our pretrained CAE (solely trained on the coarsely sampled training data), to compress this system evolution to an eight-DOF latent space. Following this, our previously trained GP is tasked with sampling at the intermediate points in time, which correspond to this finely sampled testing data. Note that the GP is trained only with the coarse data set as well. Thus, this assessment represents interpolation in \textcolor{black}{the space of varying initial conditions (which are formulated as independent parameters provided to the GP) as well as in time}. The results for the GP forecast in this assessment are shown in Figure \ref{FR_Time}. A good agreement can be observed between the true latent-space trajectory and the GP interpolated counterpart. We also direct our attention to the forecast behavior for latent dimensions \numlist{2; 6; 7} where uncertainties are seen to oscillate corresponding to the coarsely sampled training points. \textcolor{black}{The oscillation of the confidence intervals is related to the proximity to the nearest sample point in time: the trained GP exhibits greater confidence as it gets close to a value of time where data (albeit from other initial conditions) has been sampled}. We qualitatively assess the accuracy of the temporal interpolation as shown in Figure \ref{FR_Viz}, where the reconstruction from the true latent-space trajectory and the GP interpolation are compared against the truth. A good agreement is seen. Naturally, the Root Mean Square Error (RMSE) and MAE for this finely sampled test data set is seen to be higher that the case for interpolating solely on the coarser sample locations (see Table \ref{Table_Metrics_Test} for comparison), but this adds a useful utility to this ROM strategy.

\sisetup{scientific-notation = true}
\begin{table}[]
\centering
\caption{Metrics for the temporal super-resolution of testing simulations. A CAE and GP combination trained on coarsely sampled data in time is utilized to reconstruct at finer intervals}
\begin{tabular}{|c|c|c|}
\hline
Metric                          & MAE                 & RMSE                 \\ \hline
GP Interpolation+reconstruction & \num{0.0175}        & \num{0.00094}        \\ \hline
Reconstruction                  & \num{0.0084}        & \num{0.00022}        \\ \hline
\end{tabular}
\end{table}
\sisetup{scientific-notation = false}

\section{Conclusions}

The development of parameteric non-intrusive reduced-order models for systems \textcolor{black}{governed by advective partial differential equations} has great applications for cost reductions in large numerical simulation campaigns across multiple domain sciences. This article addresses their limitations associated with reduced interpretability by proposing the use of \textcolor{black}{Gaussian processes (GPs)}, conditioned on time and system control parameters that provide quantification of uncertainty. This is particularly useful when nonlinear compression techniques, such as autoencoders, are used for efficient \textcolor{black}{degree-of-freedom} reduction. In addition to the ability to interpolate \textcolor{black}{for varying initial conditions, which are formulated as independent parameters affecting the GP forecast,} we also investigate the ability of the proposed framework to interpolate in time. This addresses the fact that the sampling of a dynamical system for training these \textcolor{black}{reduced-order models} may not match the emulation temporal resolution requirement. We also remark that the modular nature of the compression and time evolution allows for the use of conventional reduced-basis methods such as the \textcolor{black}{proper orthogonal decomposition} for dynamics which are intrinsically low-dimensional.

Our results indicate that the proposed model-order reduction technique is successful at dealing with advective dynamics through assessments on the inviscid shallow-water equations. We establish this by testing on unseen initial conditions for our system evolution where a low-dimensional evolution successfully replicates high-dimensional results when coupled with a convolutional or variational autoencoder. The non-intrusive nature of our framework also allows for construction of emulators from remote-sensing or experimental data. This is of value when the underlying governing \textcolor{black}{partial differential equations} are not known a priori. Extensions to the present study shall investigate the integration of a feedback loop to sample points in control parameter space with the knowledge of prediction uncertainty. Through this, we aim to establish continually-learning model-order reduction frameworks for advective problems spanning large physical regimes. \textcolor{black}{An addition that must be explored is the ability to conserve integral variants in the reduced-order model when using the deep learning compression frameworks. This will require the utilization of modified optimization statements (with physics-informed loss functions) or intelligent application of network constraints for exact satisfaction of constraints. Our ongoing work shall address these questions.}

\section{Acknowledgements}

RM acknowledges support from the Margaret Butler Fellowship at the Argonne Leadership Computing Facility. TB, LRM, IP acknowledge support from Wave 1 of The UKRI Strategic Priorities Fund under the EPSRC Grant EP/T001569/1, particularly the \emph{Digital Twins for Complex Engineering Systems} theme within that grant and The Alan Turing Institute. IP acknowledges funding from the Imperial College Research Fellowship scheme. This material is based upon work supported by the U.S. Department of Energy (DOE), Office of Science, Office of Advanced Scientific Computing Research, under Contract DE-AC02-06CH11357. This research was funded in part and used resources of the Argonne Leadership Computing Facility, which is a DOE Office of Science User Facility supported under Contract DE-AC02-06CH11357.  This paper describes objective technical results and analysis. Any subjective views or opinions that might be expressed in the paper do not necessarily represent the views of the U.S. DOE or the United States Government. Declaration of Interests – None.

\bibliographystyle{elsart-num}
\bibliography{references}

\end{document}